\documentclass[pra,preprint,superscriptaddress,showpacs,floatfix]{revtex4}

\usepackage{graphics}           
\usepackage{bm}                 
\usepackage{amsmath}            
\usepackage{amssymb}
\usepackage{verbatim}           
\usepackage{amsthm}             
\theoremstyle{plain}            

\numberwithin{equation}{section}

\def\bra#1{{\langle#1|}}
\def\ket#1{{|#1\rangle}}
\def\bracket#1#2{{\langle#1|#2\rangle}}
\def\inner#1#2{{\langle#1|#2\rangle}}
\def\expect#1{{\langle#1\rangle}}
\def\e{{\rm e}}
\def\proj{{\hat{\cal P}}}
\def\tr{{\rm Tr}}
\def\H{{\hat H}}
\def\Hdag{{\hat H}^\dagger}
\def\Lop{{\cal L}}
\def\Ehat{{\hat E}}
\def\Edag{{\hat E}^\dagger}
\def\Shat{\hat{S}}
\def\Sdag{{\hat S}^\dagger}
\def\Ahat{{\hat A}}
\def\Adag{{\hat A}^\dagger}
\def\U{{\hat U}}
\def\Udag{{\hat U}^\dagger}
\def\Zhat{{\hat Z}}
\def\Phat{{\hat P}}

\def\id{{\hat I}}
\def\x{{\hat x}}

\def\Pr{\proj_{R}}
\def\Pl{\proj_{L}}

\begin{document}

\title{Quantum Walks driven by many coins}

\author{Todd A. Brun}
\email{tbrun@ias.edu}
\affiliation{Institute for Advanced Study, Einstein Drive, Princeton, 
              NJ 08540}

\author{Hilary A. Carteret}
\email{hcartere@cacr.math.uwaterloo.ca}
\affiliation{Department of Combinatorics and Optimization, 
             University of Waterloo, Waterloo, Ontario, N2L 3G1, Canada}

\author{Andris Ambainis}
\email{ambainis@ias.edu}
\affiliation{Institute for Advanced Study, Einstein Drive, Princeton, 
              NJ 08540}

\date{2002}

\begin{abstract}
Quantum random walks have been much studied recently, largely due
to their highly nonclassical behavior.  In this paper, we study one
possible route to classical behavior for the discrete
quantum random walk on the line:  the use of multiple quantum ``coins''
in order to diminish the effects of interference between paths.
We find solutions to this system in terms of the single coin random
walk, and compare the asymptotic limit of these solutions to numerical
simulations.  We find exact analytical expressions for the time-dependence
of the first two moments, and show that in the long time limit the
``quantum mechanical'' behavior of the one-coin walk persists.  We further
show that this is generic for a very broad class of possible walks,
and that this behavior disappears only in the limit of a new coin for
every step of the walk.
\end{abstract}



\maketitle   


\section{Introduction}

In the classical discrete random walk, a particle is located at one of
a set of positions.
In response to a random event (such as the flipping
of a coin), the particle moves in one direction or another.  This
sequence is repeated, and the motion of the particle is analyzed
statistically.  These systems provide good models for diffusion
and other stochastic processes.

Considerable work has been done recently on quantum random walks.
Quantum walks differ from classical walks in that their evolution is unitary 
and therefore reversible.  Two classes of models have been studied:
continuous \cite{FarhiGutmann98,Childs01,Childs02} and discrete
\cite{Aharonov93,Meyer96,NayaknV,Aharonov00,Ambainis01,Meyer01,Moore01,QW:higherdim,Kempe02,Konno02,Konno02b,Konno02c,Travaglione02,Du02,Yamasaki02,Sanders02,Duer02,Bach02,Kendon02},
random walks.  This paper is concerned solely with the discrete case.

A na{\"\i}ve generalization of the
classical random walk behaves in a rather boring way: if its evolution 
is unitary, it can only move in one direction and keep on going that way  
\cite{Meyer96}.  To produce nontrivial behavior, we introduce an extra
``coin'' degree of freedom (usually a single quantum bit) into the system.
Just as in the classical random walk, the outcome of a ``coin flip''
determines which way the particle moves; but in the quantum case,
both the ``flip'' of the coin and the conditional motion of the particle
are represented by unitary transformations; thus, there can be interference
between different classical paths.

This paper is concerned with quantum walks on the infinite line.  The 
particle is initially located at position $x=0$ and is free to travel off to 
infinity in either direction.   We will primarily consider the
{\it position} of the particle after some number of flips $t$.  At this
point, we assume that the position is measured.  We will look at both
the probability distribution $p(x,t) = |\inner{x}{\psi(t)}|^2$, and at
the long-time behavior of the moments $\expect\x$ and
$\expect{\x^2}$ as functions of $t$.

Unitary random walks behave quite
differently from classical random walks.  For a classical walk,
$p(x,t)$ has approximately the form of a Gaussian distribution (actually,
a binomial distribution), with a width which spreads like $\sqrt{t}$;
the variance ${\bar{x^2}}-{\bar x}^2$ grows linearly with time.
The variance in the quantum walk, by contrast, grows {\it quadratically}
with time; and the distribution $p(x,t)$ has a complicated, oscillatory
form.  Both of these are effects of interference between the
possible paths of the particle.

It should be possible to recover the classical behavior as some kind
of limit of the quantum system.  There are two obvious ways to regain
the classical result.  If the quantum coin is measured at every step,
then the record of the measurement outcomes singles out a particular
classical path.  By averaging over all possible measurement records,
one recovers the usual classical behavior \cite{QW:higherdim}.

Alternatively, rather than re-using the same coin every time, one could
replace it with a {\it new} quantum coin for each flip.  After a time
$t$ one would have accumulated $t$ coins, all of them entangled with
the position of the particle.  By measuring them, one could reconstruct
an unique classical path; averaging over the outcomes would once again
produce the classical result.  It is actually unnecessary, however, to
measure the coins.  Simply by tracing them out, one leaves the particle
in the mixed state
\begin{equation}
\rho = \sum_x p(x,t)\ket{x}\bra{x} \;,
\end{equation}
where $p(x,t)$ is the probability distribution obtained after $t$ steps
of a classical random walk.

It is therefore certainly possible to recover the classical limit.  But
these two approaches give two different routes from quantum to classical.
We might increase the number of coins used to generate the walk, cycling
among $M$ different coins, in the limit using a new coin at each step.
Or we might {\it weakly} measure the coin after each step, reaching the
classical limit with strong, projective measurements.  This is equivalent
to having the coin decohere with time.

In this paper we look at quantum random walks with multiple coins.
(The case with a decoherent coin is considered elsewhere
\cite{Letter,DecoherentCoin};
systems with decoherence of the particle have also been considered
\cite{Duer02,Kendon02,Kendon02b}.)
We begin by reviewing the approaches to solving the
single-coin walk.  We extend these to walks with multiple coins, and
derive expressions for the amplitudes of an $M$-coin walk which we can
evaluate numerically.  We look at the long-time behavior of the moments
of position both numerically and analytically, and compare it to the
usual classical random walk.

\section{The one-coin quantum walk}

Since our results
will utilize the amplitudes for the {\it single-coin} quantum walk,
we first go quickly over the single-coin derivation.  We then make use
of this in the following section to write down exact expressions for
the multi-coin case.

There are two main approaches to analyzing quantum walks.  One is to use 
Fourier analysis, which gives a good qualitative 
insight into the behavior of the system, but gives solutions
for the wavefunction in terms of some rather unpleasant integrals.
Indeed, no exact solution to them is known, and so they must be
approximated in the limit as $t \to \infty.$  The approximation 
methods become progressively more impractical as the number of coins
increases; we examine these asymptotic expressions in Appendix B.

The other approach employs combinatorics.
This gives expressions which are opaque, but are exact for 
all times.  However, evaluating these expressions exactly is prohibitive 
for long times, and so approximation methods are again required for the 
long time limit.

\subsection{Fourier Analysis of Quantum Walks}

Following the analysis in \cite{NayaknV} we consider the Hadamard walk, 
in which each coin performs the evolution
\begin{eqnarray}
\ket{\rm{R}} &\mapsto& \H\ket{\rm{R}}
  = \frac{1}{\sqrt{2}} \left(|\rm{R}\rangle 
     + |\rm{L}\rangle \right) \nonumber\\
\ket{\rm{L}} &\mapsto& \H\ket{\rm{L}}
  = \frac{1}{\sqrt{2}} \left(|\rm{R}\rangle 
                         - |\rm{L}\rangle \right)
\label{hadamard}
\end{eqnarray}
at each time step for which that coin is active, where $\rm{R}$ and $\rm{L}$
can be respectively thought of as the ``heads'' and ``tails'' states of 
the coin, or equivalently as an internal chirality state of the particle. 
The value of the coin controls the direction in which the particle moves.
When the coins shows ``R'' the particle moves right; when it shows ``L''
the particle moves left.

Let $\{\ket{x}\}$ be the position states of the particle, where the values
$x$ are integers.  We define a unitary shift operator
\begin{equation}
\Shat\ket{x} = \ket{x+1} \;, \ \ \Shat^{-1}\ket{x} = \Sdag\ket{x} = \ket{x-1} \;.
\label{shift}
\end{equation}
If $\Pr = \ket{R}\bra{R}$ and $\Pl = \ket{L}\bra{L}$ are
projectors onto the two states of the coin, then one step of the quantum
walks is given by the unitary operator
\begin{equation}
\Ehat = (\Shat\otimes\Pr + \Sdag\otimes\Pl)(\id\otimes\H) \;.
\label{evolutionOp}
\end{equation}
If the initial state of the position and coin is $\ket{\Psi_0}$, then
after $t$ steps of the walk the state is
\begin{equation}
\ket{\Psi(t)} = \Ehat^t \ket{\Psi_0} \;.
\end{equation}
For the purposes of this paper, we will usually assume that the coin
starts in the state $\ket{R}$.

Consider the wave function of the position of the particle.  For a one coin 
walk, this wavefunction has just two amplitude components, which are 
labeled by their chiralities:
\begin{equation}
 \Psi(x,t) = 
 \begin{pmatrix} a_R(x,t) \\
                 a_L(x,t)
 \end{pmatrix}.
\end{equation}
In Dirac notation this is
\begin{equation}
\ket{\Psi(t)} = \sum_x \ket{x}\otimes
  \left( a_R(x,t) \ket{R}
  + a_L(x,t) \ket{L} \right) \;.
\end{equation}

We would like to solve for $a_{L,R}(x,t)$ explicitly.  We can do
this by diagonalizing the operator $\Ehat$.
Following the analysis in \cite{NayaknV}, we use
the spatial discrete Fourier transformation, 
$\tilde{a}_{L,R}(k,t)$ for $k \in [-\pi,\pi]$
\begin{eqnarray}
\ket{\Psi(t)} &=&
  \int_{-\pi}^{\pi} \ket{k} \otimes \left( \tilde{a}_R(k,t) \ket{R} +
  \tilde{a}_L(k,t) \ket{L}\right) \, \frac{dk}{2\pi}
  \;, \nonumber\\
\tilde{a}_{L,R}(k,t) &=&
  \sum_{x=-\infty}^{\infty} a_{L,R}(x,t) e^{-ikx} \;,  \nonumber\\
\ket{k} &=& \sum_{x=-\infty}^{\infty} \ket{x} e^{ikx} \;,
\label{Fourier}
\end{eqnarray}
where the usual integral is replaced by a sum, as the variable $x$ is 
discrete.  We note that these ``momentum'' states $\ket{k}$ are eigenstates
of the shift operator $\Shat$ and its inverse:
\begin{equation}
\Shat\ket{k} = e^{-ik}\ket{k} \;, \ \ 
  \Sdag\ket{k} = e^{ik}\ket{k} \;.
\end{equation}
These states are not normalizable; but we can think of them as the
limit of the case where the number of position states is large but
finite.

In this momentum basis, the effect of the evolution
is given by the matrix
\begin{equation}
  \H_k  = \frac{1}{\sqrt{2}}
 \begin{pmatrix} e^{-ik} & e^{-ik} \\
                 e^{ik}  & -e^{ik}
 \end{pmatrix} \;.
\label{hadamardk}
\end{equation}
Since this is just a 2-by-2 matrix, it is easily diagonalized, and the
final state found in terms of its eigenvectors and eigenvalues.
The method of solution is now straightforward in principle, though
complicated in detail.  One first represents the initial state in
the $\{\ket{k}\}$ basis; applies the diagonal form of the matrix
$H_k$ $t$ times; and re-expresses the result in the original
$\{\ket{x}\}$ and $\ket{R},\ket{L}$ bases.

Nayak and Vishwanath \cite{NayaknV} carried out this program and got
exact results.  For an initial state $\ket{\Psi_0} = \ket0\ket{R}$,
\begin{eqnarray}
a_L(x,t) &=& \frac{1+(-1)^{x+t}}{2} \int_{-\pi}^{\pi} \frac{dk}{2\pi}
  \left(1 + \frac{\cos k}{\sqrt{1+\cos^2 k}} \right)
  e^{-i(\omega_k t + k x)} \;, \nonumber\\
a_R(x,t) &=& \frac{1+(-1)^{x+t}}{2} \int_{-\pi}^{\pi} \frac{dk}{2\pi}
  \frac{e^{ik}}{\sqrt{1+\cos^2 k}}
  e^{-i(\omega_k t + k x)} \;.
\label{onecoinexact}
\end{eqnarray}
In the exponents, we define the frequencies $\omega_k$ by
$\sin\omega_k = \sqrt{1/2}\sin k$, taking $\omega_k \in [-\pi/2,\pi/2]$.

These integrals, while exact, are difficult to evaluate.
Nayak and Vishwanath found an approximation to these results,
valid in the limit of large $t$, by making an asymptotic expansion of
the integrals, to get the approximate expressions
\begin{eqnarray}
a_L(x,t) &\approx&
  \frac{1 + (-1)^{x+t} }{\sqrt{8\pi t(1-x^2/t^2)\sqrt{1-2x^2/t^2}} }
  \nonumber\\
&& \times (1-x/t)\left( \exp i(t \phi(x/t)+\pi/4) +
  {\rm c.c.} \right) \;, \nonumber\\
a_R(x,t) &\approx&
  \frac{1 + (-1)^{x+t} }{\sqrt{8\pi t(1-x^2/t^2)\sqrt{1-2x^2/t^2}} } 
  \nonumber\\
&& \times \left((x/t+i\sqrt{1-2x^2/t^2}) \exp i(t \phi(x/t)+\pi/4)
  + {\rm c.c.} \right) \;.
\label{onecoinapprox}
\end{eqnarray}
The results are conveniently represented in terms of a variable
$\alpha=x/t$.  This approximation is valid only within the range
$-\sqrt{1/2} < \alpha < \sqrt{1/2}$; outside this range the amplitude
becomes negligibly small.   The phase $\phi(\alpha)$ is given by
\begin{equation}
\phi(\alpha) = \arcsin \sqrt{ \frac{1-2\alpha^2}{2-2\alpha^2} }
  + \alpha \arcsin \sqrt{ \frac{1-2\alpha^2}{1-1\alpha^2} } \;.
\end{equation}

\subsection{Combinatorial Analysis}

We can do a completely different analysis of the one-coin walk,
using a combinatorial rather than a Fourier argument, similar to that
done in \cite{Meyer96} and \cite{Ambainis01}.
Suppose we start with $x=0$ and the coin in state $\ket{R}$.
We see how the state evolves after the first few flips of the coin:
\begin{eqnarray}
\ket{\Psi_0} &=& \ket0\ket{R} \;, \nonumber\\
\ket{\Psi(1)} &=& \sqrt{1/2} \left( \ket1\ket{R}
  + \ket{-1}\ket{L} \right) \;, \nonumber\\
\ket{\Psi(2)} &=& \sqrt{1/2^2} \left( \ket2\ket{R}
  + \ket0\ket{L} 
  + \ket0\ket{R}
  - \ket{-2}\ket{L} \right) \;.
\end{eqnarray}
At time $t$ there will be $2^t$ terms, each with amplitude
$\pm \sqrt{1/2^t}$.  Each of these terms corresponds to a
possible path for a {\it classical} random walk.  To get the
amplitude for a particular position $x$, conditioned on the coin showing
R or L, one must sum up the amplitudes for all the paths which
end at the position $x$ with the appropriate coin face showing.
This amounts to a nice problem in combinatorics.

For the coin to reach position $x$ at time $t$, it must have moved to
the left a total of $N_L = (t-x)/2$ times, and to the right a total
of $N_R = (t+x)/2$ times.  For a term corresponding to a given classical
path, it turns out that the phase depends only on the values of $N_L$ and
$N_R$, and the number $C$ of {\it clusters} of consecutive L flips.
The details of the derivation are given in Appendix A, so we just
quote the answer here:
\begin{align}\label{combas1}
 a_L(x,t) &= \frac{1}{\sqrt{2^t}} \left[ \sum_{C=1}^{``N"} (-1)^{N_L-C}
            \binom{N_L-1}{C-1} \binom{N_R}{C-1} \right], \nonumber\\
 a_R(x,t) &= \frac{1}{\sqrt{2^t}} \left[ \sum_{C=1}^{``N"} (-1)^{N_L-C}
            \binom{N_L-1}{C-1} \binom{N_R}{C} \right],
\end{align}
where the summation is to $N_L$ for $x\geq 0$ and to $N_R+1$ for $x<0$
(but only for $N_L \neq 0$; for $N_L=0$ the amplitude is always $2^{-t/2}$).
We can, of course, re-express $N_R$ and $N_L$ as functions of $x$ and $t.$

If we had started instead with the coin in the state $\ket{L}$,
the amplitudes would then become
\begin{align}\label{combbs1}
 b_L(x,t) &= \frac{1}{\sqrt{2^t}} \left[ \sum_{C=1}^{``N"} (-1)^{N_L-C}
            \binom{N_L-1}{C-1} \binom{N_R}{C-1} \frac{N_R - 2C +2}{N_R}
            \right], \nonumber\\ 
 b_R(x,t) &= \frac{1}{\sqrt{2^t}} \left[ \sum_{C=1}^{``N"} (-1)^{N_L-C}
            \binom{N_L-1}{C-1} \binom{N_R}{C} \frac{N_R-2C}{N_R}
            \right], 
\end{align}
where the summation is to $N_L$ for $x\geq 0$ and to $N_R+1$ for $x<0$,
and $N_L=0$ is done separately, as before.

The solution given by Eq.~(\ref{combas1}) looks very different from
that given by Eq.~(\ref{onecoinexact}); it is rather extraordinary
that they should describe exactly the same outcome.  Moreover, a
long-time approximation similar to Eq.~(\ref{onecoinapprox})
can be derived from Eq.~(\ref{combas1}), by using the asymptotics
of Jacobi polynomials.  See \cite{Ambainis01} for details.

If the coin starts in an arbitrary state 
\begin{equation}
 \psi_{\rm{init}} = \alpha|R\rangle + \beta|L\rangle 
\end{equation}
the amplitudes will be 
\begin{align}
 g_L(x,t) &= \alpha a_L(x,t) + \beta b_L(x,t) \\
 g_R(x,t) &= \alpha a_L(x,t) + \beta b_L(x,t),
\end{align}
respectively.   

\section{Walks with multiple coins}

For a walk driven by $M$ coins, the wavefunction has $2^M$ components; for 
example, a two coin walk will be described by the wavefunction
\begin{equation}
 |\Psi(x,t) \rangle = 
 \begin{pmatrix} a_{RR}(x,t) \\
                 a_{RL}(x,t) \\
                 a_{LR}(x,t) \\
                 a_{LL}(x,t)
 \end{pmatrix}.
\end{equation}
At each step, the particle moves in the direction dictated by the coin that is 
active at that step, with the other coins remaining inert until it is their 
turn once again.  We will initially assume that we cycle through the coins 
in a deterministic, regular way, although we will discover later that we can 
relax that condition in certain specific ways.

The unitary transformation which results from flipping the $m$th coin is
\begin{equation}
\Ehat_m = ( \Shat\otimes\proj_{{R}m} + \Sdag\otimes\proj_{{L}m} )
  (\id \otimes \H_m) \;,
\end{equation}
where $\H_m$ is the Hadamard transformation on the $m$th coin, and
$\proj_{{L,R}m}$ is the projector onto the $m$th coin being in state
L or R, respectively.  If we cycle among the coins, doing a total of
$t$ flips ($t/M$ flips with each coin), then the state will be
\begin{equation}
\ket{\Psi(Mt)} = (\Ehat_M \cdots \Ehat_1)^{t/M} \ket{\Psi_0} \;.
\end{equation}

We have simulated this system numerically for different numbers $M$
of coins.  The probability distributions $p(x,t)$ agree with the
classical results only up to $t=M$; beyond that, they diverge sharply,
with the multicoin distributions exhibiting highly oscillatory behavior
and rapid spreading (linear with $t$), similar to the behavior of the
single-coin case.  (See figure 1.)

\begin{figure}[t]
\includegraphics{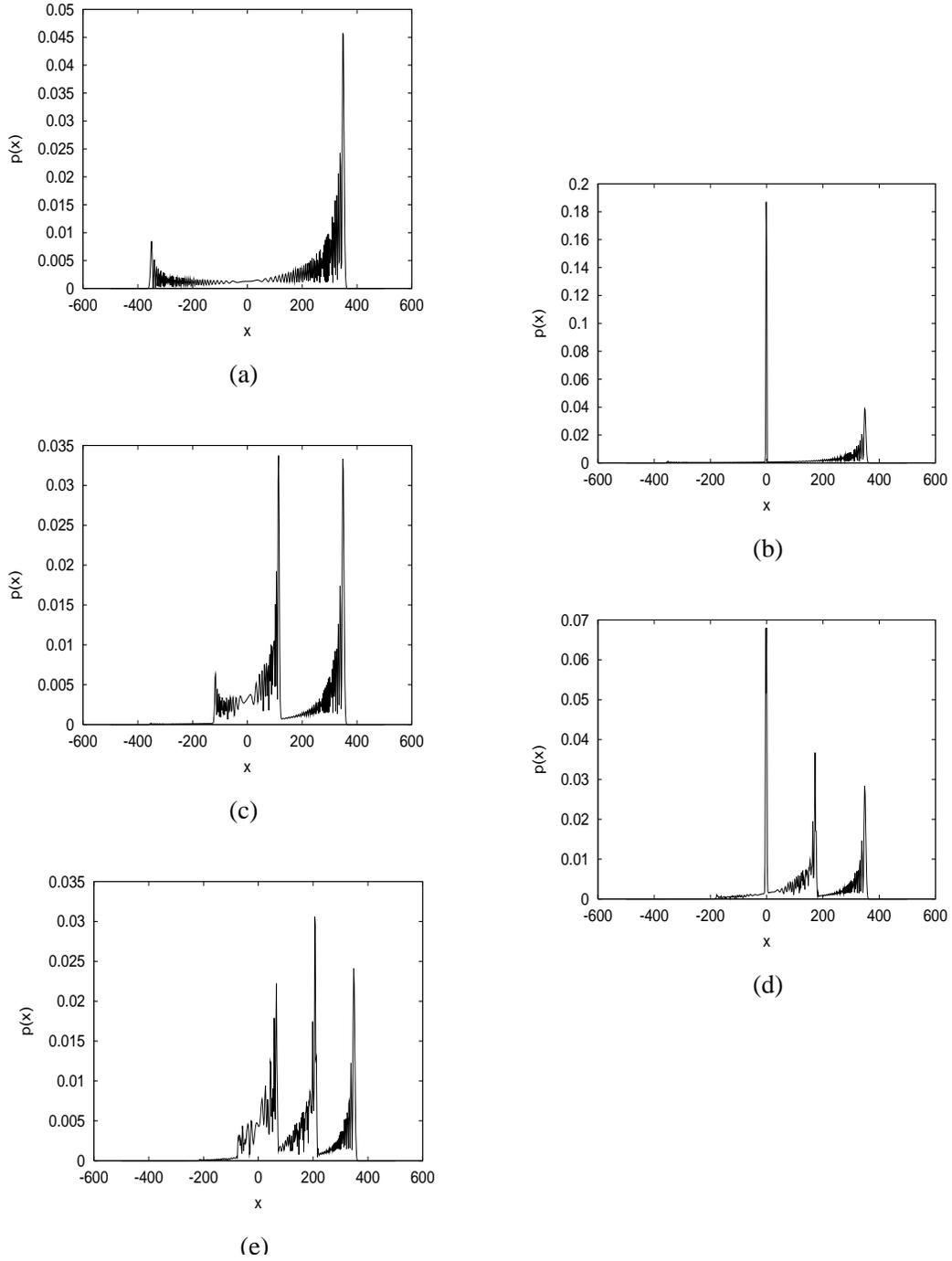}
\caption{\label{fig1} These figures plot the particle position
distribution $p(x)$ vs. $x$ at time $t=500$ for multicoin quantum
random walks with 1--5 coins.  In each of these plots, all the
coins begin in state $\ket{R}$, which is the reason for the
asymmetry of the distributions.  Note that only even points $x$ are
plotted, since $p(x)=0$ for odd values of $t-x$.}
\end{figure}

How can we understand these $M$-coin results?
An important thing to notice is that these $M$ unitary transformations
$\{\Ehat_m\}$ all commute:
\begin{equation}
[\Ehat_m,\Ehat_n] = 0 \;.
\end{equation}
Because of that, the order in which they are listed is irrelevant.  We
can equally well write
\begin{equation}
\ket{\Psi(Mt)} = (\Ehat_M)^{t/M} \cdots (\Ehat_1)^{t/M} \ket{\Psi_0} \;.
\end{equation}
We can either let the walk 
evolve for $t$ steps using $M$ coins, or we could use one coin for $t/M$ 
steps, and then replace the coin with another one, and use that for 
another $t/M$ steps, and so on, until we've used all $M$ coins.  The 
final probability distribution will be the same. 

Suppose we decide to use our coins cyclically.
Then it can be seen that at the end of the walk, the $M$ coins act 
as a quantum memory of the last $M$ moves that the particle made.  The 
``which path'' information from before then is lost, so we will see 
interference between paths arising from moves dating from before 
that time.

Now consider $M$ walks controlled by one coin each,
concatenated in the following recipe: Do the first one-coin walk.  Measure
the final state of the coin only, doing nothing that would give
any further information about the position of the particle.
Reset the coin state to the appropriate starting
state for the next walk, and use the conditioned state produced by
the measurement as the starting state for the next one-coin walk.
Repeat this procedure until you've concatenated $M$ walks in this way. 
The \emph{final} probability distribution obtained after the last coin has 
been measured will be exactly the same as the on obtained after measuring 
the $M$ coins from the single cyclic $M$ coin walk.

How can we use this to solve the multicoin walk?
Assume (for the moment) that the coins all start in the state $\ket{R}$,
and the particle starts in the state $\ket0$.
We flip the coins $t/M$ times each (with $t/M$ even for convenience).
After each coin has been flipped $t/M$ times, we measure it, and find it
in state L or R.  We then define {\it conditional evolution operators}
for the particle position:
\begin{eqnarray}
 \hat{A}_L(t) &=& \sum_{j=-t/2M}^{t/2M-1} a_L(2j,t/M)\hat{S}^{2j} \nonumber\\
 \hat{A}_R(t) &=& \sum_{j=-t/2M+1}^{t/2M} a_R(2j,t/M)\hat{S}^{2j},
\label{conditionalevolve}
\end{eqnarray}
where $x=2j.$  If we flipped $M$ coins $t/M$ times each and then measured them,
getting $N$ results R and $M-N$ results L, then the particle must
have ended up in the (unnormalized) state
\begin{equation}
\left[\hat{A}_R(t)\right]^N\left[\hat{A}_L(t)\right]^{M-N} \ket0 \;.
\end{equation}
The probability 
of the particle arriving at the final position $x$ at time $Mt$ is then
$p(x,Mt)= |\Psi(x,Mt)|^2,$ which is 
\begin{equation}
 p(x,Mt)= \sum_{N=0}^{M} \binom{M}{N} | \bra{x}
  \left[\hat{A}_R(t)\right]^N\left[\hat{A}_L(t)\right]^{M-N} \ket0 |^2.
\label{multicoinprobs}
\end{equation}
Note that this distribution doesn't depend on the order in which the
coins were flipped, nor the order in which the measurement results
occurred, provided that the measurement on coin $m$ was made after it
had been flipped exactly $t/M$ times.

For the amplitudes $a_{L,R}(x,t)$ in (\ref{conditionalevolve}), we can
use any of the expressions (\ref{onecoinexact}), (\ref{onecoinapprox}),
or (\ref{combas1}), whichever is convenient.  We can also calculate
them numerically; Eq.~(\ref{multicoinprobs}) provides a much more
practical method of numerical calculation for multicoin systems than
direct simulation, which requires a Hilbert space that grows exponentially
in dimension with the number of coins.  By contrast, the difficulty
of evaluating (\ref{multicoinprobs}) grows only linearly with the number
of coins.  We have used this to simulate systems with large numbers of
coins, which would otherwise be impractical; see, e.g., figure 2.

\begin{figure}[t]
\includegraphics{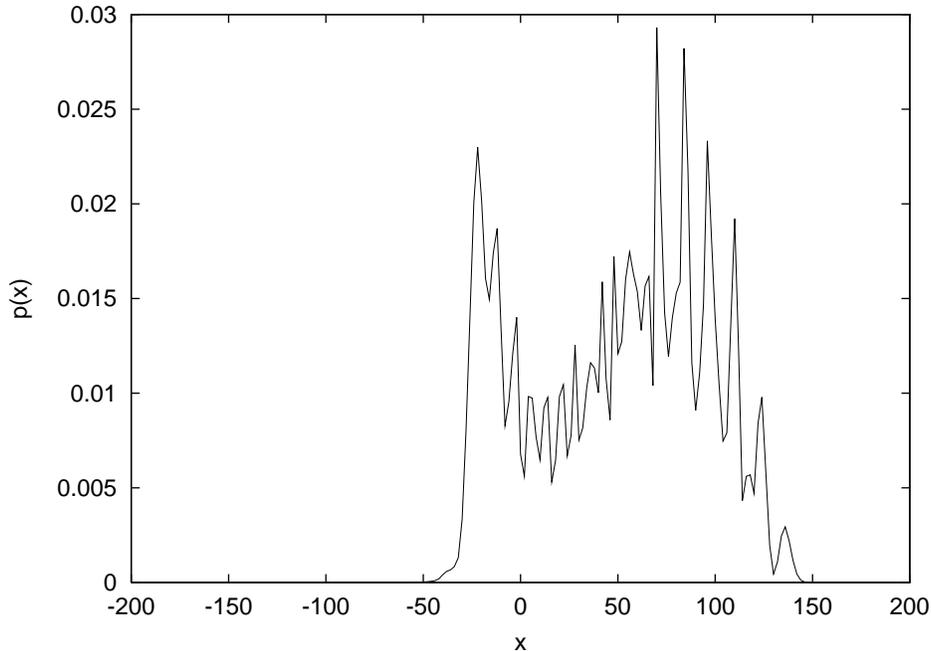}
\caption{\label{fig2} This figure plots the numerical results for
$p(x,t)$ at $t=200$ for 20 coins, each initially in the state $\ket{R}$,
using the formula (\ref{multicoinprobs}).  As we see, the highly
oscillatory form of the probability distribution persists even for
large numbers of coins.  (Only even points are plotted.)}
\end{figure}

One can find asymptotic expressions for $p(x,t)$ at long times which
are analogous to the single-coin expressions (\ref{onecoinapprox}),
though the difficulties mount rapidly with the number of coins.
We plot the approximate solution for two coins in figure 3, which matches
the numerical results of figure 1b quite well.  For details of
the calculation, see Appendix B.

\begin{figure*}[t]
\includegraphics{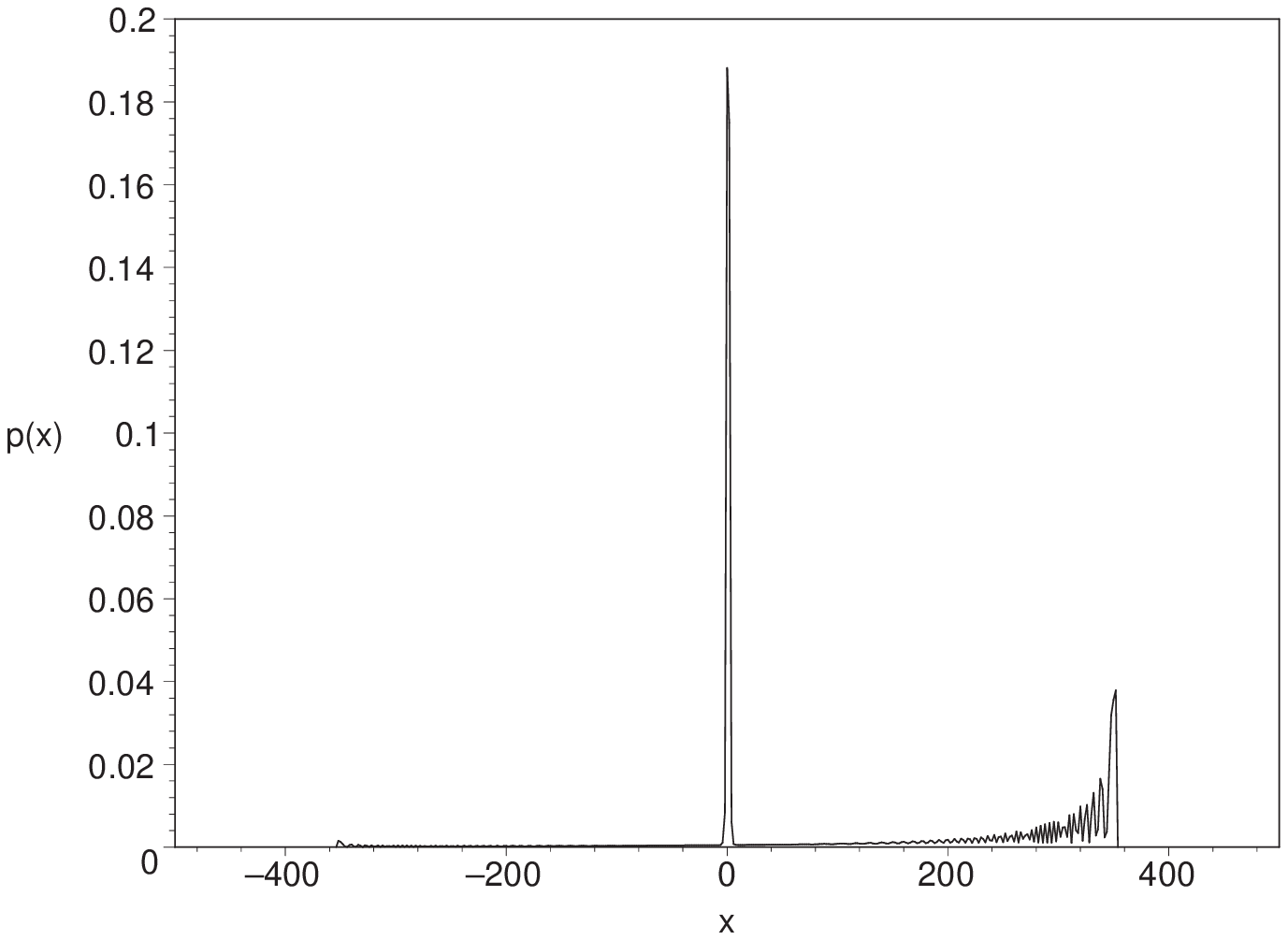}
\caption{\label{fig3} This figure plots $p(x,t)$ at $t=500$ for the
two-coin quantum walk, using the long-time approximation derived
in Appendix B.  Comparison to figure 1b shows that this result matches
direct numerical simulation very closely.}
\end{figure*}

While having points of similarity, these multicoin walks
differ from the tensor product walks studied in
\cite{QW:higherdim}, in which each component in the tensor product 
corresponds to a different spatial dimension.  There are also points
of similarity to quantum Parrondo games, as studied in \cite{Meyer01}.

\section{Moments of the distributions}

While the probability distributions $p(x,t)$ for the quantum random
walk clearly differ markedly from their classical counterparts, it is
difficult to use this to give a quantitative criterion for quantum
vs. classical behavior.  In particular, we would like to know
if the quantum walks become ``more classical'' in any meaningful sense
as we let $M\rightarrow t$.  It would be useful to look at one or
two numbers instead of the entire probability distribution in answering
this question.

One obvious criterion suggested by the single-coin case is to look at
the long-time behavior of the moments of $p(x,t)$, and see if (for instance)
the variance increases linearly or quadratically with time.  Since
it becomes more and more difficult to calculate $p(x,t)$ as we increase
$M$, this might seem like an unhelpful approach; but as it happens,
we can extract expressions for the long time behavior of the moments
without having to evaluate $p(x,t)$ itself \cite{Letter}.

Note that in the following two subsections, we fix the number of
coins $M$ (or more generally the dimension $D$ of the ``coin'') and
then go to the long-time limit.  What constitutes a long time will
depend on the choice of $M$ or $D$.  In the final subsection, we let the
number of coins used (or the dimension of the ``coin'') increase
with time, and look at the long-time limit in that case.

\subsection{The general case}

Let's consider a very general linear random walk, which includes the
multicoin case as a subcase.  Let the ``coin'' be a $D$-dimensional
system with an initial state $\ket{\Phi_0}$; let $\proj_{R,L}$ be two
orthogonal projectors on the Hilbert space of the coin, such that
$\Pr + \Pl = \id$, and $\tr\Pr = \tr\Pl = D/2$.
We also define a unitary transformation $\U$
which ``flips'' the coin.  Then one step of the quantum random walk is
given by the unitary operator
\begin{equation}
\Ehat \equiv \left(\Shat \otimes \Pr + \Sdag \otimes \Pl\right)
  \left( \id \otimes \U \right) \;,
\end{equation}
where $\Shat,\Sdag$ are the usual shift operators (\ref{shift})
on the particle position.
The full initial state of the system (particle and coin) is
\begin{equation}
\ket{\Psi_0} = \ket0 \otimes \ket{\Phi_0} \;.
\end{equation}

We can identify the eigenvectors $\ket{k}$ of $\Shat,\Sdag$,
with eigenvalues $\exp(\mp ik)$, as defined in (\ref{Fourier}).
In particular,
\begin{equation}
\ket{0} = \int_{-\pi}^{\pi} \frac{dk}{2\pi} \ket{k} \;.
\end{equation}
In the $k$ basis, the evolution operator becomes
\begin{eqnarray}
\Ehat\left( \ket{k} \otimes \ket\Phi \right)
  &=& \ket{k} \otimes \left( \e^{-ik} \Pr + \e^{ik} \Pl\right) \U
  \ket\Phi \;, \nonumber\\
  &\equiv& \ket{k} \otimes \U_k \ket\Phi \;,
\end{eqnarray}
where $\U_k$ is a unitary operator on the coin degree of freedom.

Let the quantum random walk proceed for $t$ steps.  Then the state
evolves to
\begin{eqnarray}
\ket{\Psi_0} = \ket0 \otimes \ket{\Phi_0} &\rightarrow&
  \Ehat^t \ket{\Psi_0} \nonumber\\
&& = \int_{-\pi}^{\pi} \frac{dk}{2\pi}
  \ket{k} \otimes \left(\U_k\right)^t \ket{\Phi_0} \equiv \ket{\Psi(t)} \;.
\end{eqnarray}
The probability to reach a point $x$ at time $t$ is
\begin{eqnarray}
p(x,t) &=& \bra{\Psi(t)} \left( \ket{x}\bra{x}\otimes \id \right)
  \ket{\Psi(t)} \nonumber\\
&=& \frac{1}{(2\pi)^2} \int dk \int dk' \bracket{k}{x}\bracket{x}{k'}
  \bra{\Phi_0} \left(\Udag_k\right)^t \left(\U_{k'}\right)^t \ket{\Phi_0}
  \nonumber\\
&=& \frac{1}{(2\pi)^2} \int dk \int dk' \e^{-ix(k-k')}
  \bra{\Phi_0} \left(\Udag_k\right)^t \left(\U_{k'}\right)^t \ket{\Phi_0} \;.
\end{eqnarray}

This will, in general, be difficult to calculate.  However, all we are
interested in are the {\it moments} of this distribution:
\begin{eqnarray}
\expect{\x^m}_t &=& \frac{1}{(2\pi)^2} \sum_x x^m p(x,t) \nonumber\\ 
&=& \frac{1}{(2\pi)^2} \sum_x x^m \int dk \int dk' \e^{-ix(k-k')}
  \bra{\Phi_0} \left(\Udag_k\right)^t \left(\U_{k'}\right)^t \ket{\Phi_0} \;.
\end{eqnarray}
We can then invert the order of operations and do the $x$ sum first.
This sum can be exactly carried out in terms of derivatives of
the delta function:
\begin{equation}
\frac{1}{2\pi} \sum_x x^m \e^{-ix(k-k')}
  = (-i)^m \delta^{(m)}(k-k') \;.
\end{equation}
Inserting this result back into our expression for $\expect{x^m}_t$ yields
\begin{equation}
\expect{\x^m}_t =
  \frac{(-i)^m}{2\pi} \int dk \int dk' \delta^{(m)}(k-k')
  \bra{\Phi_0} \left(\Udag_k\right)^t \left(\U_{k'}\right)^t \ket{\Phi_0} \;.
\end{equation}
We can then integrate this by parts to obtain expressions such as
\begin{eqnarray}
\expect{\x^m}_t &=&
  \frac{(-i)^m}{2\pi} \int dk \int dk' \delta^{(m)}(k-k')
  \bra{\Phi_0} \left(\Udag_k\right)^t \left(\U_{k'}\right)^t \ket{\Phi_0}
  \nonumber\\
&=& \frac{(-i)^m}{2\pi} \int dk \int dk' \delta(k-k')
  \bra{\Phi_0} \left(\Udag_k\right)^t 
  \left[ \frac{d^m}{{dk'}^m} \left(\U_{k'}\right)^t \right]\ket{\Phi_0}
  \nonumber\\
&=& \frac{(-i)^m}{2\pi} \int dk
  \bra{\Phi_0} \left(\Udag_k\right)^t 
  \left[ \frac{d^m}{dk^m} \left(\U_k\right)^t \right]\ket{\Phi_0}
\end{eqnarray}

For the first moment we get
\begin{equation}
\expect{\x}_t = - \frac{i}{2\pi} \int dk
  \bra{\Phi_0} \left(\Udag_k\right)^t \left[ \frac{d}{dk}
  \left(\U_k\right)^t \right] \ket{\Phi_0} \;,
\label{firstmoment1}
\end{equation}
where
\begin{eqnarray}
\frac{d\U_k}{dk} &=&
  \left( - i \e^{-ik}\Pr + i \e^{ik}\Pl \right) \U =
  - i (\Pr - \Pl) \U_k \equiv - i \Zhat \U_k \nonumber\\
\frac{d\Udag_k}{dk} &=&
  i \Udag_k (\Pr - \Pl) \equiv i \Udag_k \Zhat \nonumber\\
\Zhat &\equiv& \Pr - \Pl = \id - 2 \Pl \;.
\label{zdef}
\end{eqnarray}
Substituting this into (\ref{firstmoment1}) gives us
\begin{equation}
\expect{\x}_t = - \frac{1}{2\pi} \sum_{j=1}^t \int dk
  \bra{\Phi_0} \left(\Udag_k\right)^j \Zhat
  \left(\U_k\right)^j \ket{\Phi_0} \;.
\label{firstmoment2}
\end{equation}

We can carry out a similar integration by parts to get the second moment:
\begin{eqnarray}
\expect{\x^2}_t &=&
  \frac{1}{2\pi} \int dk
  \bra{\Phi_0} \left[ \frac{d}{dk} \left(\Udag_k\right)^t \right]
  \left[ \frac{d}{dk} \left(\U_k\right)^t \right] \ket{\Phi_0} \nonumber\\
&=& \frac{1}{2\pi} \sum_{j=1}^t \sum_{j'=1}^t \int dk
  \bra{\Phi_0} \left(\Udag_k\right)^j \Zhat \left(\U_k\right)^{j-j'}
  \Zhat \left(\U_k\right)^{j'} \ket{\Phi_0} \;.
\label{secondmoment1}
\end{eqnarray}

Can we now evaluate these expressions?  Let us suppose that we can find
the eigenvectors $\ket{\phi_{kl}}$ and corresponding eigenvalues
$\exp(i\theta_{kl})$ of $\U_k$.  We expand the initial state
\begin{equation}
\ket{\Phi_0} = \sum_l c_{kl} \ket{\phi_{kl}} \;.
\end{equation}
After $t$ steps
\begin{equation}
\left(\U_k\right)^t\ket{\Phi_0}
  = \sum_l c_{kl} \ket{\phi_{kl}} \e^{i\theta_{kl}t} \;.
\end{equation}
Substituting this into the equation for the first moment
(\ref{firstmoment2}) we get
\begin{equation}
\expect{x}_t = -t + \frac{1}{\pi} \int dk \sum_{l,l'}
  c_{kl}^* c_{kl'} \bra{\phi_{kl}} \Pl \ket{\phi_{kl'}} 
  \sum_{j=1}^t \e^{i(\theta_{kl'}-\theta_{kl})j} \;.
\label{firstmoment3}
\end{equation}
If the unitary matrix is nondegenerate, then most of the terms
in (\ref{firstmoment3}) will be {\it oscillatory}; hence, over time,
they will average to zero.  Only the diagonal terms in the above sum
are nonoscillatory.  We can therefore write
\begin{eqnarray}
\expect{x}_t &=& C_1 t + {\rm oscillatory\ terms} \;, \nonumber\\
C_1 &=& -1 + \frac{1}{\pi} \int dk \sum_l
  |c_{kl}|^2 \bra{\phi_{kl}} \Pl \ket{\phi_{kl}} \;.
\label{firstmoment4}
\end{eqnarray}

Making the same substitutions in the equation for the second moment
(\ref{secondmoment1}), we get
\begin{equation}
\expect{x^2}_t = \frac{1}{2\pi} \int dk \sum_{l,l',l''}
  c_{kl}^* c_{kl'} \bra{\phi_{kl}} \Zhat \ket{\phi_{kl''}} 
  \bra{\phi_{kl''}} \Zhat \ket{\phi_{kl'}} 
  \sum_{j,j'=1}^t \e^{i(\theta_{kl''}-\theta_{kl})j}
  \e^{i(\theta_{kl'}-\theta_{kl''})j'} \;.
\label{secondmoment2}
\end{equation}
Once again, most of these terms are oscillatory.  There are two sets
of nonoscillatory terms:  terms with $l=l'=l''$ (which give a quadratic
dependence on $t$) and terms with $j=j'$ and $l=l'$ (which give a linear
dependence).  Therefore we can write the second moment as
\begin{eqnarray}
\expect{x^2}_t &=& C_2 t^2 + {\rm oscillatory\ terms} + O(t) \;, \nonumber\\
C_2 &=& 1 - \frac{2}{\pi} \int dk \sum_l
  |c_{kl}|^2 \bra{\phi_{kl}} \Pl \ket{\phi_{kl}}
  \bra{\phi_{kl}} \Pr \ket{\phi_{kl}} \;.
\label{secondmoment3}
\end{eqnarray}
Of course, if the spectrum of $\U_k$ is degenerate, we will have to
modify (\ref{secondmoment3}) and (\ref{firstmoment4}) to include appropriate
cross terms.  This does not, however, alter the qualitative behavior.
We see that generically in the long time limit, the first moment of
the quantum random walk on the line will undergo a linear drift, and the
variance will grow quadratically with time,
so long as the coin is a finite-dimensional system.

\subsection{The multicoin model}

Now let us specialize to the case of our multicoin model.  In this case
our coin is a tensor-product of $M$ 2-level coins, with a Hilbert
space of $2^M$ dimensions.  The ``flip operator'' is
\begin{equation}
\U = \left[ \H \otimes \id^{\otimes M-1} \right] \Phat \;,
\end{equation}
where $\H$ is the usual Hadamard operator (\ref{hadamard})
and $\Phat$ is a cyclic permutation of the $M$ coins:
\begin{equation}
\Phat ( \ket{\psi_0} \otimes \ket{\psi_1} \otimes \cdots
  \otimes \ket{\psi_{M-1}} ) =
\ket{\psi_1} \otimes \cdots \otimes
  \ket{\psi_{M-1}} \otimes \ket{\psi_0} ) \;.
\end{equation}
The two projectors onto the flip results are
\begin{eqnarray}
\Pr &=& \ket{R}\bra{R} \otimes \id^{\otimes M-1} \;, \nonumber\\
\Pl &=& \ket{L}\bra{L} \otimes \id^{\otimes M-1} \;.
\end{eqnarray}
When we switch to the $k$ representation, these give us the effective
unitary evolution
\begin{equation}
\U_k = \left[ \H_k \otimes \id^{\otimes M-1} \right] \Phat \;,
\end{equation}
where $\H_k$ is given by (\ref{hadamardk}).
In the case $M=1$, $\U_k = \H_k$.

The eigenvectors of $\H_k$ are
\begin{equation}
\ket{\pm} = \frac{1}{\sqrt{2}} \left( 1 + \cos^2 k
  \pm \cos k \sqrt{1 + \cos^2 k} \right)^{-1/2}
  \begin{pmatrix} \e^{-ik} \\
      \mp \sqrt{2}\e^{\pm i\omega_k}-\e^{-ik}
  \end{pmatrix} \;,
\label{hkeigenvectors}
\end{equation}
with eigenvalues $\e^{i(\pi+\omega_k)}$, $\e^{-i\omega_k}$, respectively,
where we define $\omega_k$ to satisfy
\begin{equation}
\sin\omega_k \equiv \frac{1}{\sqrt{2}}\sin k \;, \ \ 
-\pi/2 < \omega_k < \pi/2 \;.
\label{hkeigenvalues}
\end{equation}

A basis for the full $M$-coin space is given by product vectors of
$\ket{+}$ and $\ket{-}$.  (E.g., for $M=2$ the basis vectors would be
$\ket{--}, \ket{-+}, \ket{+-}, \ket{++}$.)  Can we find the eigenvectors
and eigenvalues of $\U_k$ in terms of this basis?

Note first that if we apply $\U_k$ to such a basis vector, we get another
basis vector multiplied by a phase.  Second, if we apply $\U_k$ to a
basis vector $M$ times, we get the {\it same} basis vector back, multiplied
by a phase.

Two obvious eigenvectors appear at once:  the vectors
\begin{eqnarray}
\ket{++ \cdots +} &\equiv& \ket{+}^{\otimes M} \;, \nonumber\\
\ket{-- \cdots -} &\equiv& \ket{-}^{\otimes M} \;,
\end{eqnarray}
are both eigenvectors of $\U_k$ with eigenvalues
$\e^{i(\pi+\omega_k)}$, $\e^{-i\omega_k}$, respectively.  Let us suppose
for the moment that $M$ is a prime number.  Then for all basis vectors
$\ket\phi$ other than those two, the vectors
\begin{equation}
\left(\U_k\right)^j \ket\phi \;,\ \ 0 \le j \le M-1
\end{equation}
are all distinct.  Together they span an $M$-dimensional subspace which
is preserved under the action of $\U_k$.  If $\ket\phi$ contains exactly
$m$ $\ket{-}$s and $(M-m)$ $\ket{+}$s, then every vector in this subspace
is an eigenvector of $(\U_k)^M$ with eigenvalue
$\exp i((M-m)(\pi+\omega_k)-m\omega_k)$.  This implies that the subspace
is spanned by $M$ eigenvectors of $\U_k$ with eigenvalues
\begin{equation}
\lambda_n = \e^{i(-m\omega_k+(M-m)(\pi+\omega_k)+2\pi n)/M} \;,\ \ 
  0 \le n < M \;.
\label{eigenvalue}
\end{equation}

Let's restrict ourselves to this $M$-dimensional subspace for the moment.
Starting with our original basis state $\ket\phi$, we label the basis
states which span this space
\begin{equation}
\ket{\phi_j} = \Phat^j\ket\phi \;.
\end{equation}
The eigenvectors of $\U_k$ must have the form
\begin{equation}
\ket{\chi_n} = \frac{1}{\sqrt{M}} \sum_{j=0}^{M-1}
  e^{i\nu_{nj}} \ket{\phi_j} \;.
\end{equation}
Plugging the above expression and equation (\ref{eigenvalue}) into the equation
\begin{equation}
\U_k\ket{\chi_n} = \lambda_n \ket{\chi_n}
\end{equation}
gives the result
\begin{equation}
\nu_{nj+1} = \nu_{nj}
  - \left(m\omega_k - (M-m)(\pi+\omega_k) - 2\pi n\right)/M
  + \theta_{j+1} \;,
\end{equation}
where we define $\theta_j$ to be $\e^{-i\omega_k}$ if the first factor
of $\ket{\phi_j}$ is $\ket{-}$ and to be $\e^{i(\pi+\omega_k)}$ if
the first factor is $\ket{+}$.  Since the overall phase is arbitrary,
we take $\nu_0 = 0$, which gives us
\begin{equation}
\nu_{nj} =
  - \frac{j}{M} \left(m\omega_k - (M-m)(\pi+\omega_k) + 2\pi n\right)
  + \sum_{j'=1}^j \theta_{j'} \;.
\label{eigenvector}
\end{equation}

For each value of $m$, $1 \le m \le M-1$, there are $\frac{1}{M}{M\choose m}$
such $M$-dimensional subspaces which are preserved by the action of $\U_k$;
each eigenvalue is thus $\frac{1}{M}{M\choose m}$-fold degenerate.
However, this doesn't complicate the expressions (\ref{firstmoment4})
and (\ref{secondmoment3}), because the cross-terms vanish:
\begin{equation}
\bra{\phi_{kl}} \Pr \ket{\phi_{kl'}} =
\bra{\phi_{kl}} \Pl \ket{\phi_{kl'}} = 0
\end{equation}
for all such degenerate eigenvectors $\ket{\phi_{kl}},\ket{\phi_{kl'}}$.
Thus, we can use the expressions we've already derived.

Using the results in Eq.~(\ref{eigenvector}) we can calculate the necessary
matrix elements:
\begin{eqnarray}
\bra{\chi_n} \Pl \ket{\chi_n} &=&
  \frac{1}{M}\left[ m \bracket{-}{L}\bracket{L}{-}
  + (M-m) \bracket{+}{L}\bracket{L}{+} \right] \nonumber\\
&=&  \frac{1}{M}\Biggl[ \frac{ m (\sqrt{1+\cos^2 k} - \cos k)^2 }{2
  (1+\cos^2 k - \cos k \sqrt{1+\cos^2 k})} \nonumber\\
&&\ \  + \frac{ (M-m) (\sqrt{1+\cos^2 k} + \cos k)^2 }{2
  (1+\cos^2 k + \cos k \sqrt{1+\cos^2 k})} \Biggr] \;, \nonumber\\
\bra{\chi_n} \Pr \ket{\chi_n} &=&
  \frac{1}{M}\left[ m \bracket{-}{R}\bracket{R}{-}
 + (M-m) \bracket{+}{R}\bracket{R}{+} \right] \nonumber\\
&=&  \frac{1}{M}\Biggl[ \frac{ m }{2
  (1+\cos^2 k - \cos k \sqrt{1+\cos^2 k})} \nonumber\\
&&\ \  + \frac{ M-m }{2
  (1+\cos^2 k + \cos k \sqrt{1+\cos^2 k})} \Biggr] \;.
\end{eqnarray}

We also need the amplitudes of the coin's initial state.  If
\begin{equation}
\ket{\Phi_0} = \ket{\psi_0} \otimes \cdots \otimes \ket{\psi_0}
  = \ket{\psi_0}^{\otimes M} \;,
\end{equation}
then
\begin{equation}
|\bracket{\Phi_0}{\chi_n}|^2 = |\bracket{\psi_0}{-}|^{2m}
  |\bracket{\psi_0}{+}|^{2(M-m)}
  \sum_{j,j'} \frac{1}{M} \e^{i(\nu_{nj}-\nu_{nj'})} \;.
\end{equation}
Note that
\begin{equation}
\sum_{n=0}^{M-1} \e^{i(\nu_{nj}-\nu_{nj'})} = M \delta_{jj'} \;.
\end{equation}
We now have everything we need to get the moments!

To evaluate the expression (\ref{firstmoment4}) to get $C_1$ we calculate
\begin{eqnarray}
\sum_l |c_{kl}|^2 \bra{\phi_{kl}}\Pl\ket{\phi_{kl}}
&=& \sum_{m=1}^{M-1} \frac{1}{M} {M\choose m} \sum_{n=0}^{M-1}
  |\bracket{\Phi_0}{\phi_n}|^2
  \bra{\phi_n}\Pl\ket{\phi_n} \nonumber\\
&& + |\bracket{\psi_0}{-}|^2 \bracket{-}{L}\bracket{L}{-}
  + |\bracket{\psi_0}{+}|^2 \bracket{+}{L}\bracket{L}{+} \nonumber\\
&=& \sum_{m=0}^M {M\choose m} |\bracket{\psi_0}{-}|^{2m}
  |\bracket{\psi_0}{+}|^{2(M-m)} \nonumber\\
&&\ \ \  \times  \frac{1}{M}\Biggl[ \frac{ m (\sqrt{1+\cos^2 k} - \cos k)^2 }{2
  (1+\cos^2 k - \cos k \sqrt{1+\cos^2 k})} \nonumber\\
&&\ \ \ \ \  + \frac{ (M-m) (\sqrt{1+\cos^2 k} + \cos k)^2 }{2
  (1+\cos^2 k + \cos k \sqrt{1+\cos^2 k})} \Biggr] \;.
\label{c1expression}
\end{eqnarray}

We can simplify this considerably by noting that this equation
(\ref{c1expression}) has the form
\begin{equation}
\sum_{m=0}^M {M\choose m} p^m (1-p)^{M-m} [ (m/M) (A-B) + B ]
  = p A + (1-p) B \;,
\end{equation}
where
\begin{eqnarray}
p &=& |\bracket{\psi_0}{-}|^2 \;, \nonumber\\
A &=& \frac{ (\sqrt{1+\cos^2 k} - \cos k)^2 }{2
  (1+\cos^2 k - \cos k \sqrt{1+\cos^2 k})} \;, \nonumber\\
B &=& \frac{ (\sqrt{1+\cos^2 k} + \cos k)^2 }{2
  (1+\cos^2 k + \cos k \sqrt{1+\cos^2 k})} \;,
\end{eqnarray}
and we've invoked some combinatorial identities
\begin{eqnarray}
\sum_{m=0}^M {M\choose m} p^m (1-p)^{M-m}
  &=& \left( p + (1-p) \right)^M = 1 \;, \nonumber\\
\sum_{m=0}^M {M\choose m} p^m (1-p)^{M-m} (m/M)
  &=& p \sum_{m=0}^{M-1} {M-1\choose m} p^m (1-p)^{M-m-1}
  = p \;.
\end{eqnarray}
This leaves us with the simple expression
\begin{eqnarray}
\sum_l |c_{kl}|^2 \bra{\phi_{kl}}\Pl\ket{\phi_{kl}} &=&  \Biggl[
  \frac{ |\bracket{\psi_0}{-}|^2 (\sqrt{1+\cos^2 k} - \cos k)^2 }{2
  (1+\cos^2 k - \cos k \sqrt{1+\cos^2 k})} \nonumber\\
&&\ \  + \frac{ |\bracket{\psi_0}{+}|^2 (\sqrt{1+\cos^2 k} + \cos k)^2 }{2
  (1+\cos^2 k + \cos k \sqrt{1+\cos^2 k})} \Biggr] \;,
\label{c1expression2}
\end{eqnarray}
in which all $M$ dependence is gone!

Let us choose the initial condition $\ket{\psi_0} = \ket{R}$.  Then
\begin{eqnarray}
|\bracket{\psi_0}{-}|^2
&=& \frac{1}{2 (1+\cos^2 k - \cos k \sqrt{1+\cos^2 k})} \nonumber\\
|\bracket{\psi_0}{+}|^2
&=& \frac{1}{2 (1+\cos^2 k + \cos k \sqrt{1+\cos^2 k})} \;.
\end{eqnarray}
We can plug this into the above equation (\ref{c1expression2}); after
a bit of algebra, this boils down to the result
\begin{equation}
\sum_l |c_{kl}|^2 \bra{\phi_{kl}}\Pl\ket{\phi_{kl}}
  = \frac{1}{2(1 + \cos^2 k)} \;,
\end{equation}
This gives us the final equation for $C_1$,
\begin{equation}
C_1 = -1 + \frac{1}{2\pi} \int \frac{dk}{1+\cos^2 k}
  = -1 + 1/\sqrt{2} \;,
\label{c1expression3}
\end{equation}
which exactly matches the observed numerical results.  (See figure 4.)

\begin{figure}[t]
\includegraphics{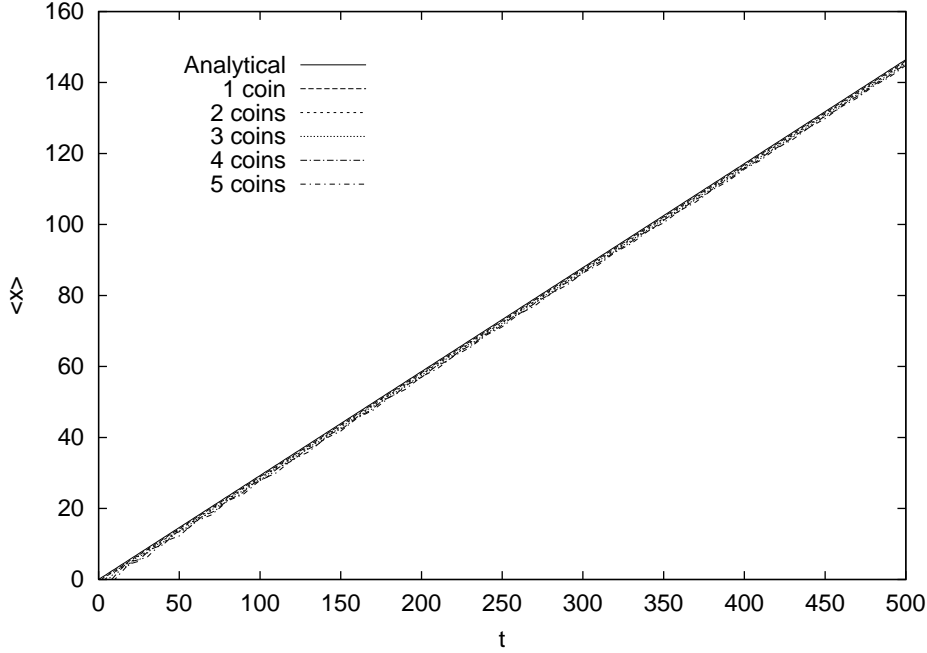}
\caption{\label{fig4} This plots the first moment of position vs. time
for multicoin quantum random walks with $M=$1--5 coins, using both
direct numerical simulation and the analytical result (\ref{c1expression3}).
As calculated, the long time behavior is essentially independent of
the number of coins, and grows linearly with time.  All coins begin in
the initial state $\ket{R}$.}
\end{figure}

We can do a similar derivation for the second moment, in order to
get $C_2$.  For this we use
\begin{eqnarray}
&& \sum_l |c_{kl}|^2 \bra{\phi_{kl}}\Pl\ket{\phi_{kl}}
  \bra{\phi_{kl}}\Pr\ket{\phi_{kl}}\ \ \ \ \ \nonumber\\
&& = \sum_{m=0}^M {M\choose m} |\bracket{\psi_0}{-}|^{2m}
  |\bracket{\psi_0}{+}|^{2(M-m)} \nonumber\\
&&\ \ \  \times  \frac{1}{M}\Biggl[ \frac{ m (\sqrt{1+\cos^2 k} - \cos k)^2 }{2
  (1+\cos^2 k - \cos k \sqrt{1+\cos^2 k})} \nonumber\\
&&\ \ \ \ \  + \frac{ (M-m) (\sqrt{1+\cos^2 k} + \cos k)^2 }{2
  (1+\cos^2 k + \cos k \sqrt{1+\cos^2 k})} \Biggr] \nonumber\\
&&\ \ \  \times  \frac{1}{M}\Biggl[ \frac{ m }{2
  (1+\cos^2 k - \cos k \sqrt{1+\cos^2 k})} \nonumber\\
&&\ \ \ \ \  + \frac{ (M-m) }{2
  (1+\cos^2 k + \cos k \sqrt{1+\cos^2 k})} \Biggr] \;.
\label{c2expression}
\end{eqnarray}

We can again simplify by observing that (\ref{c2expression})
has the form
\begin{eqnarray}
\sum_{m=0}^M {M\choose m} p^m (1-p)^{M-m}
  [ (m/M) (A-B) + B ]
  [ A + (m/M) (B-A) ] \nonumber\\
= A B + p(1-p) \frac{M-1}{M} (A-B)^2 \;.
\end{eqnarray}
If we specialize once more to $\ket{\psi_0} = \ket0$, we can make
similar algebraic simplifications to get 
\begin{eqnarray}
C_2 &=& 1 - \frac{1}{2\pi} \int dk \left[
  \frac{1+2\cos^2 k}{(1+\cos^2 k)^2}
  - \frac{1}{M} \frac{\cos^2 k}{(1+\cos^2 k)^2} \right] \nonumber\\
&=& 1 - \frac{5}{4\sqrt{2}} + \frac{1}{M} \frac{1}{4\sqrt{2}} \;.
\end{eqnarray}

Of course, numerically we calculated not the second moment but the
{\it variance} of $x$.  Given our results,
\begin{equation}
\expect{\x^2}_t - {\expect{\x}_t}^2 = (C_2-C_1^2) t^2 + O(t)
  + {\rm osc.\ terms} \;,
\end{equation}
we get the result
\begin{equation}
C_2 - C_1^2 = \frac{3 - 2\sqrt{2} + 1/M}{4\sqrt{2}} \;.
\label{variance}
\end{equation}
Note that this doesn't vanish as $M$ becomes large!

The derivation used above was only strictly valid for prime $M$.  However,
it turns out that the additional complications arising from composite $M$
all cancel out, yielding the same expressions for $C_1$ and $C_2$.
Upon comparison to the numerical results, we see that these analytical
expressions for the first moment and the variance match
the simulations extremely well.  (See figure 5.)

\begin{figure}[t]
\includegraphics{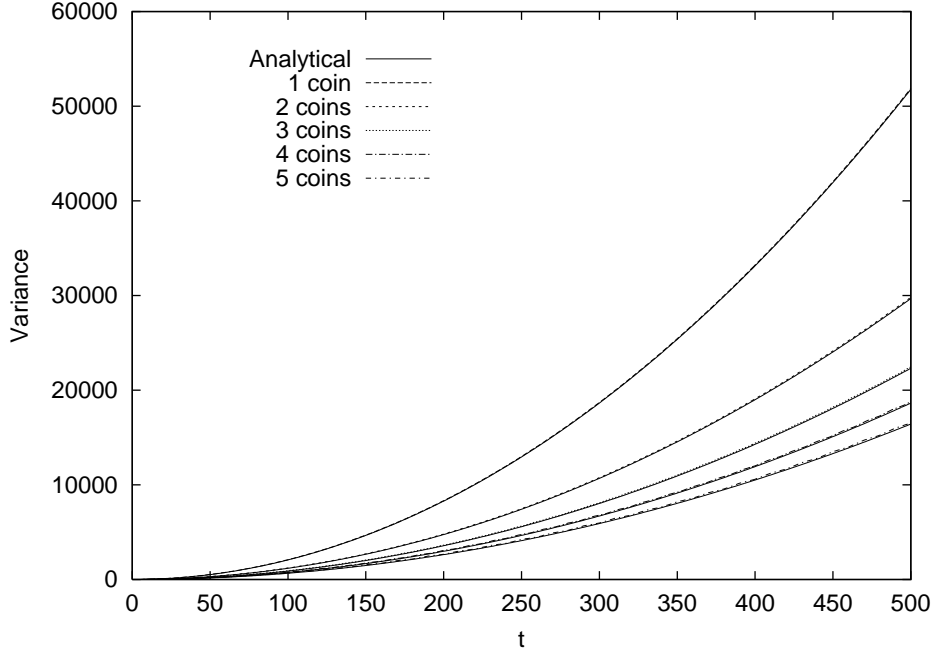}
\caption{\label{fig5} This plots the variance of position vs. time
for multicoin quantum random walks with $M=$1--5 coins, using both
direct numerical simulation and the analytical result (\ref{variance}).
All coins begin in the initial state $\ket{R}$.}
\end{figure}

\subsection{Constant flips per coin}

We see that in the long time limit, an $M$-coin quantum random walk with
fixed $M$ has qualitative behavior similar to that one of the one-coin
walk, and is markedly nonclassical.  A reasonable question to ask is,
how long is a long time?  Suppose we flip each coin at most $d$ times, and
add more coins as we go to long times, so that $M$ and $t$ are both growing
in a fixed ratio $t/M \rightarrow d$.  Does this still behave nonclassically?
The fact that $C_2 - C_1^2$ does not vanish as $M\rightarrow\infty$
makes this conjecture plausible.  If so, how big must $d$ be for
this nonclassical behavior to manifest itself?

Let us now consider a quantum random walk on a line driven by a sequence
of two-level coins, each flipped $d$ times.  By the reordering principle
described in section III, this is the same as having a single coin which
is measured and reset to the initial state $\ket{\Phi_0}$ after every $d$
steps.  We describe this evolution by a {\it superoperator} $\Lop$ which
acts on the density matrix of the particle and coin:
\begin{eqnarray}
\rho \rightarrow \rho' &=& \Lop\rho \nonumber\\
&=& \sum_n \Ahat_n \left(\Ehat\right)^d \rho \left(\Edag\right)^d \Adag_n \;,
\end{eqnarray}
where $\Ehat$ is the usual unitary evolution (\ref{evolutionOp}), and
\begin{eqnarray}
\Ahat_0 &=& \id \otimes \ket{\Phi_0}\bra{R} \;, \nonumber\\
\Ahat_1 &=& \id \otimes \ket{\Phi_0}\bra{L} \;.
\end{eqnarray}

The superoperator $\Lop$ represents $d$ steps of the walk.  Note that
\begin{equation}
\Lop\rho
= \tr_{\rm coin} \left\{ \left(\Ehat\right)^d \rho \left(\Edag\right)^d \right\}
  \otimes \ket{\Phi_0}\bra{\Phi_0} \;.
\end{equation}

We rewrite this in terms of the eigenvectors $\ket{k}$ of $\Shat,\Sdag$.
In the $k$ basis, a general density operator for the joint
particle/coin system is written
\begin{equation}
\rho = \int \frac{dk}{2\pi}\int \frac{dk'}{2\pi} \ket{k}\bra{k'} \otimes
  \chi_{kk'} \;,
\end{equation}
and the evolution superoperator becomes
\begin{eqnarray}
\Lop\rho
  &=& \int \frac{dk}{2\pi}\int \frac{dk'}{2\pi} \ket{k}\bra{k'} \otimes
  \sum_{n=R,L} \ket{\Phi_0}\bra{n} \left( \H_k \chi_{kk'} \Hdag_{k'} \right)
  \ket{n}\bra{\Phi_0} \nonumber\\
&\equiv& \int \frac{dk}{2\pi}\int \frac{dk'}{2\pi} \ket{k}\bra{k'} \otimes
  \Lop_{kk'} \chi_{kk'} \;,
\end{eqnarray}
where $\Lop_{kk'}$ is now a superoperator on the coin degree of freedom
alone.

The initial state of the system (particle and coin) is
\begin{eqnarray}
\rho_0 &=& \ket{\Psi_0}\bra{\Psi_0} = \ket0\bra0
  \otimes \ket{\Phi_0}\bra{\Phi_0} \nonumber\\
&=& \int \frac{dk}{2\pi}\int \frac{dk'}{2\pi} \ket{k}\bra{k'} \otimes
  \ket{\Phi_0}\bra{\Phi_0} \;.
\end{eqnarray}
Let the quantum random walk proceed for $t$ steps.  Then the state
evolves to
\begin{equation}
\rho_t = 
  \int \frac{dk}{2\pi}\int \frac{dk'}{2\pi} \ket{k}\bra{k'} \otimes
  \Lop_{kk'}^{t/d}  \ket{\Phi_0}\bra{\Phi_0} \;.
\end{equation}
The probability to reach a point $x$ at time $t$ is
\begin{eqnarray}
p(x,t) &=& \tr\left\{ (\ket{x}\bra{x}\otimes\id) \rho_t \right\}
  \nonumber\\
&=& \frac{1}{(2\pi)^2} \int dk \int dk' \bracket{k}{x}\bracket{x}{k'}
  \tr\left\{ \Lop_{kk'}^{t/d} \ket{\Phi_0}\bra{\Phi_0} \right\}
  \nonumber\\
&=& \frac{1}{(2\pi)^2} \int dk \int dk' \e^{-ix(k-k')}
  \tr\left\{ \Lop_{kk'}^{t/d} \ket{\Phi_0}\bra{\Phi_0} \right\} \;.
\end{eqnarray}
We are interested in the moments of this distribution.
\begin{eqnarray}
\expect{\x^m}_t &=& \sum_x x^m p(x,t) \nonumber\\ 
&=& \frac{1}{(2\pi)^2} \sum_x x^m \int dk \int dk' \e^{-ix(k-k')}
  \tr\left\{ \Lop_{kk'}^{t/d} \ket{\Phi_0}\bra{\Phi_0} \right\} \;.
\end{eqnarray}

Just as before, we invert the order of operations and do the $x$ sum first,
which yields
\begin{equation}
\expect{\x^m}_t =
  \frac{(-i)^m}{2\pi} \int dk \int dk' \delta^{(m)}(k-k')
  \tr\left\{ \Lop_{kk'}^{t/d} \ket{\Phi_0}\bra{\Phi_0} \right\} \;.
\end{equation}
We can then integrate this by parts.
In carrying out this integration by parts, we will need
\begin{eqnarray}
\frac{d}{dk}  \Lop_{kk'} \rho
  &=& - i \sum_n \Ahat_n 
  \left( \sum_{j=0}^{d-1} (\H_k)^j \Zhat (\Hdag_k)^j \right)
  (\H_k)^d \rho (\Hdag_k)^d \Adag_n \nonumber\\
&=& - i \ket{\Phi_0}\bra{\Phi_0} \tr\left\{ \left( \sum_{j=0}^{d-1}
  (\H_k)^j \Zhat (\Hdag_k)^j \right)
  (\H_k)^d \rho (\Hdag_k)^d \right\} \;,
\label{lopderiv}
\end{eqnarray}
where $\H_k$ is given by (\ref{hadamardk}) and
$\Zhat$ by (\ref{zdef}).

Using this, we can carry out the integration by parts for the
first moment to get
\begin{eqnarray}
\expect{\x}_t &=& - \frac{1}{2\pi} \sum_{j=0}^{t/d-1} \int dk
  \tr\left\{ \Zhat_{kd} \Lop_{kk}^j \ket{\Phi_0}\bra{\Phi_0} \right\}
  \nonumber\\
&=& - \frac{1}{2\pi} \sum_{j=0}^{t/d-1} \int dk
  \tr\left\{ \Zhat_{kd} \ket{\Phi_0}\bra{\Phi_0} \right\}
  \nonumber\\
&=& - \frac{t}{d} \int \frac{dk}{2\pi}
  \bra{\Phi_0} \Zhat_{kd} \ket{\Phi_0} \;,
\label{firstmoment}
\end{eqnarray}
where we define
\begin{equation}
\Zhat_{kd} = \sum_{l=1}^d (\Hdag_k)^l \Zhat (\H_k)^l \;.
\end{equation}
Generically, then, we see from (\ref{firstmoment}) that the first moment
of position grows linearly with time provided
$\bra{\Phi_0}\Zhat_{kd}\ket{\Phi_0}$ is nonzero.

We can carry out a similar integration by parts to get the second moment:
\begin{eqnarray}
\expect{\x^2}_t &=& \frac{1}{2\pi} \int dk \Biggl[
 \sum_{j=0}^{t/d-1}
  \tr\left\{ \left(\Zhat_{kd}\right)^2 \Lop_{kk}^j
  \ket{\Phi_0}\bra{\Phi_0} ) \right\} \nonumber\\
&& + \sum_{j=0}^{t/d-1} \sum_{j'=0}^{j-1} 
  \tr\left\{ \Zhat_{kd} \Lop_{kk}^{j-j'} ( ( \Lop_k^{j'}
  \ket{\Phi_0}\bra{\Phi_0} ) \Zhat_{kd} +
  \Zhat_{kd} \Lop_{kk}^{j'} \ket{\Phi_0}\bra{\Phi_0} ) \right\} \Biggr]
  \nonumber\\
&=& \frac{1}{2\pi} \int dk \Biggl[
  (t/d) \bra{\Phi_0} \left(\Zhat_{kd}\right)^2 \ket{\Phi_0} +
  (t^2/d^2 - t/d) \bra{\Phi_0} \Zhat_{kd} \ket{\Phi_0}^2 \;.
\label{secondmoment}
\end{eqnarray}
So here we expect the second moment to grow quadratically with time,
qualitatively like the single-coin walk.

Let's choose the initial condition $\ket{\Phi_0} = \ket{R}$, and look
at a few values of $d$.  We are interested in the coefficients
$C_1$ and $C_2$, where
\begin{eqnarray}
\expect{\x}_t &=& C_1 t \;, \nonumber\\
\expect{\x^2}_t &=& C_2 t^2 + O(t) \;.
\end{eqnarray}
From the equations (\ref{firstmoment}), (\ref{secondmoment}), we get
\begin{eqnarray}
C_1 &=& - \frac{1}{d} \int \frac{dk}{2\pi}
  \bra{R} \Zhat_{kd} \ket{R} \;, \nonumber\\
C_2 &=& \frac{1}{d^2} \int \frac{dk}{2\pi}
  \bra{R} \Zhat_{kd} \ket{R}^2 \;.
\end{eqnarray}
From the definitions of $\Zhat_{kd}$ and $\H_k$ we can work out
the values of $C_{1,2}$ in specific cases.

For $d=1$, the matrix element $\bra{R}\Zhat_{k1}\ket{R}=0$, so
both $C_1$ and $C_2$ vanish.  We recover the classical case.

For $d=2$, the matrix element $\bra{R}\Zhat_{k2}\ket{R}=\cos 2k$, so
$C_1$ vanishes in this case as well.  However, $C_2 = 1/8 \ne 0$.
So even with only two flips per coin, the variance already grows
quadratically with time.

For $d=3$, the matrix element
$\bra{R}\Zhat_{k3}\ket{R}=\cos 2k + \sin^2 2k$, so both $C_1$ and $C_2$
are nonzero.  In this case, $C_1 = -1/6$ and $C_2 = 7/72$.

If we let both $d$ and $t/d$ become large, then we can evaluate
(\ref{firstmoment}) and (\ref{secondmoment}) to get
$C_1 \rightarrow -1 + \sqrt{1/2}$ and $C_2 \rightarrow 1 - 5/\sqrt{32}$,
which are in agreement with our earlier results for large $M$ in
the multicoin case.

\section{Conclusions}

We have examined one possible path from quantum to classical behavior:
the use of multiple coins (or, more generally, higher-dimensional
systems) to drive the walk and reduce the effects of interference.
We have seen that quantum behavior, as typified by quadratic growth of
the variance with time, persists except in the extreme limit of a new
coin for every step.  Furthermore, quadratic growth of the variance
seems to be a generic feature of such unitary walks.

From the multicoin example we might speculate that classical behavior
is only recovered in the limit where the coin system retains enough
information about the walk to reconstruct a unique classical path for
the particle.  Since there are $2^t$ such paths up to time $t$, the
coin must have a Hilbert space dimension which grows exponentially
in time in order to exhibit classical behavior.

Another plausible route to classical behavior adds decoherence
to the coin, which also effectively suppresses interference effects.
The results of this study are presented elsewhere \cite{DecoherentCoin}.

\begin{acknowledgments}

We would like to thank Bob Griffiths, Lane Hughston, Viv Kendon,
Michele Mosca, and Bruce Richmond
for useful conversations.  TAB acknowledges financial support from
the Martin A.~and Helen Chooljian Membership in Natural Sciences,
and DOE Grant No.~DE-FG02-90ER40542.  AA was supported by NSF grant
CCR-9987845, and by the State of New Jersey.  HAC was supported by
MITACS, the Fields Institute, and the NSERC CRO project ``Quantum
Information and Algorithms.''

\end{acknowledgments}

\appendix

\section{Combinatorial derivation of one-coin walk}

Assume that the coin starts at $x=0$.
After $t$ coin flips, the particle might have reached locations anywhere 
between $x=-t, -t+2, \ldots, t-2,t.$  The number of paths that finish at 
a point $x$ after $t$ steps is 
\begin{equation}
 \# \;\text{paths} = \binom{t}{(t-x)/2}.
\end{equation}
Paths ending with the coin in state $|L\rangle$ cannot interfere with paths 
ending with the coin in state $|R\rangle.$

A given path can be written as a sequence $S$ of $t$ symbols, $L$ or $R$ like 
this one:
\begin{equation}
 L\;R\;R\;L\;R
 \underbrace{L\;L\;L}_{\substack{\text{a cluster}\\ \text{of}\;3\;{Ls}}}
 R\;R \ldots
\end{equation}
We let $N_L(S)$ be the number of $L$'s in $S$ and
$N_R(S)$ the number of $R$'s.  Then
\begin{align}
 &N_L(S)+N_R(S) =t, &N_R(S)-N_L(S) =x \;,
\end{align}
so we can write
\begin{align}
  &N_R(S)=(t+x)/2,  &N_L(S)=(t-x)/2 \;.
\end{align}

Each path has an associated phase factor, which can take the values $\pm 1.$
A factor of $-1$ is acquired whenever the coin flips two successive lefts, 
so a pair $LL$ contributes a phase factor of $-1,$ the sequence $LLL$ 
produces a factor of $+1,$ and $LLLL$ results in a $-1,$ and so on. 
Each cluster of $m$ $L$s contributes a factor of $(-1)^{m-1}$ to the phase 
for that path. 

Let $C(S)$ be the number of $L$-clusters in $S.$  (An $L$-cluster must 
contain at least one $L.$)  Swapping an $L$ from one cluster to another 
changes the phase contributed by each cluster by $-1,$ thus having no 
overall effect on the phase of the path as a whole.  Therefore, only the 
number of $L$-clusters and the total number of $L$s matters. 

If the number of $L$-clusters is $C$ and the total number of $L$s is $N_L,$ 
then these could be arranged thus
\begin{equation}
 \underbrace{L \quad L \quad L \quad \ldots \quad L}_{C-1\quad
 \text{singletons}}\quad
 \underbrace{L\;L\;\ldots\;L}_{\substack{N_L-(C-1) \\ \text{consecutively}}}.
\end{equation}
This path will have a phase of 
\begin{equation}
 \varphi(S) = (-1)^{N_L-(C-1)-1} = (-1)^{N_L-C}.
\end{equation}
So when $N_L-C$ is odd the overall phase is $-1.$

Each cluster must have at least one $R$ between itself and the next cluster. 
Call these groups of $R$s ``partitions.''  Every sequence $S$ is then a 
succession of alternating clusters and partitions.  Let
\begin{equation}
 P(S) = \# \; \text{of partitions in} \quad S.
\end{equation}
Clearly $C(S)$ cannot be higher than $N_L(S),$ and $P(S)$ cannot be higher 
than $N_R(S).$  Also, it must be the case that 
\begin{equation}
 P(S)=C(S) \quad \text{or} \quad C(S)\pm 1. 
\end{equation}
Allocating $N_L$ $L$s among $C$ clusters can be done in
${N_L-1\choose C-1}$ ways; similarly, $N_R$ $R$s can be allocated among
$P$ partitions in ${N_R-1\choose P-1}$ ways.  For a given $C$ and $P$,
then, the total number of paths must be
\begin{equation}
 \# \;\text{paths} \; = \binom{N_L-1}{C-1}\binom{N_R-1}{P-1} \;.
\end{equation}

Let us now fix $t$ and $x,$ and hence also $N_L$ and $N_R.$  What values 
of $C$ and $P$ are possible, and how many paths have each value?

{\bf{Case I:}} $x>0 \implies N_R > N_L.$  The number of clusters, $C,$ can 
range from $0$ (only when $x=t$) to $N_L.$  The number of partitions $P$ 
can be $C,C+1$ or $C-1,$ except when $C=0,1.$

{\bf{Case II:}} $x<0 \implies N_L>N_R.$  This time $C$ can range from $1$ to 
$N_R+1.$  Likewise, $P$ can take values of $C,C+1,C-1$ except at the extremes. 

{\bf{Case III:}} $x=0 \implies N_L=N_R$ and $C$ and $P$ can be anything from
$1$ to $N_L=N_R.$

Consider now the arrangements of clusters and partitions.  There are
four distinct possible arrangements.

\noindent Arrangement 1: $P=C-1.$ 
\begin{equation}
 \left[\text{cluster}\right]\; \text{partition} \; 
 \left[\text{cluster}\right]\; \text{partition} \ldots 
 \left[\text{cluster}\right].
\end{equation}
Arrangement 2: $P=C$, starting with a partition.  This looks like
\begin{equation}
 \text{partition} \; \left[\text{cluster}\right] \; \text{partition}\;
 \ldots \; \left[\text{cluster}\right].
\end{equation}
Arrangement 3: $P=C,$ starting with a cluster.
\begin{equation}
 \left[\text{cluster}\right]\;\text{partition}\;\left[\text{cluster}\right]
 \; \ldots \; \text{partition}.
\end{equation}
Arrangement 4: $P=C+1.$
\begin{equation}
 \text{partition}\;\left[\text{cluster}\right]\;\text{partition}\; 
 \left[\text{cluster}\right] \; \ldots \; \text{partition}.
\end{equation}

We adopt the convention
\begin{equation}
 \binom{a}{b} = 0, \quad \text{if} \quad b>a,\;b<0,\quad \text{or} \quad a<0.
\end{equation}
Then we can write that the amplitude to reach $x$ after $t$ coin flips, 
ending in an $L$ is:
\begin{equation}
 a_L(x,t) = \frac{1}{\sqrt{2^t}} \left[ \sum_{C=1}^{``N"} (-1)^{N_L-C}
            \binom{N_L-1}{C-1} 
            \left\{ \binom{N_R-1}{C-2} + \binom{N_R-1}{C-1} \right\} \right],
\end{equation}
where the summation is to $N_L$ for $x\geq 0$ and to $N_R+1$ for $x<0.$

Likewise, the amplitude to reach $x$ in $t$ flips, ending in an $R$ is 
\begin{equation}
 a_R(x,t) = \frac{1}{\sqrt{2^t}} \left[ \sum_{C=1}^{``N"} (-1)^{N_L-C}
            \binom{N_L-1}{C-1} 
            \left\{ \binom{N_R-1}{C-1} + \binom{N_R-1}{C} \right\} \right],
\end{equation}
but only if $N_L \neq 0$; that case (which corresponds to $t=x$) 
always has amplitude $2^{-t/2}$.

We can use the binomial relations
\begin{align}
 \binom{N_R-1}{C-2} + \binom{N_R-1}{C-1} &= \binom{N_R}{C-1} \nonumber\\
 \binom{N_R-1}{C-1} + \binom{N_R-1}{C} &= \binom{N_R}{C},
\end{align}
to simplify these somewhat to give the final equations (\ref{combas1}).

If we had started instead with the coin in the state $\ket{L}$, all 
sequences which begin with an $L$ (i.e., arrangements 1 and 3)
would pick up an extra minus sign.  The amplitudes then become
those given by the equations (\ref{combbs1}).

\section{Asymptotic approximation to the Fourier integrals}

Following the analysis in \cite{NayaknV}, the Fourier analysis of the 
many coin quantum walk can be completed in the same way.  
We represent the state of the particle in the $\{\ket{k}\}$ basis,
and make use of the eigenvectors and eigenvalues of $\H_k$, as
given by (\ref{hkeigenvectors}) and (\ref{hkeigenvalues}).
We will also use the identity
\begin{equation}\label{Aeig}
 \pm \sqrt{2}e^{\mp i\omega_k} - e^{-ik} = \pm \sqrt{1 + \cos^2(k)} - \cos(k)
\end{equation}
We can now use the tensor-product structure of the evolution matrix and the 
integral simplifications in \cite{NayaknV} to simplify the integrals for 
$a_R$ and $a_L.$  
The expression for $a_R$ becomes
\begin{equation}\label{foldoverR}
 \int_{-\pi}^{\pi}\left(1-\frac{\cos(k)}{\sqrt{1+\cos^2(k)}}\right)
    e^{i((\omega_k+\pi)t-kx)}dk  
= (-1)^{x+t}\int_{-\pi}^{\pi}\left(1+\frac{\cos(k')}
                                          {\sqrt{1+\cos^2(k')}}\right)
            e^{-i(\omega_{k'}t+k'x)}dk' 
\end{equation}
where $k'=k-\pi, \quad \omega_{k'} = -\omega_k$ and we are free to move the 
limits of integration because the integral is over a whole period of the 
function.  The two terms in the integral for $a_L$ can be simplified 
in a similar way:
\begin{multline}\label{foldoverL1}
 \int_{-\pi}^{\pi} \frac{-e^{ik}(\sqrt{1+\cos^2(k)}+\cos(k))}
                    {((1+\cos^2(k))+\cos(k)\sqrt{1+\cos^2(k)})}
                    e^{i(\omega_k+\pi)t-kx}dk \\
= -(-1)^{x+t}\int_{-\pi}^{\pi} \frac{e^{ik'}}{2\sqrt{1+\cos^2(k')}}
                             e^{-i(\omega_{k'}t+k'x)}dk' \;.
\end{multline}

Before we can write down a closed form for the $M$-coin momentum 
wave-function, we need one more construction.  This is a timelike parameter 
that we'll call $|\tau|$.
Let $t_i$ be the total number of turns that coin $i$ has.  The tensor product
structure of the evolution matrix for the multicoin system means that the 
eigenvalues for the combined system are just products of the eigenvalues 
for the one coin system.  So we'll see things like
\begin{align}
 \lambda_{\bf{k}}^{t} &= e^{-i\omega_kt_1}e^{i(\omega_k+\pi)t_2}
                        e^{-i\omega_kt_3}e^{i(\omega_k+\pi)t_4}
                        e^{i(\omega_k+\pi)t_5} \ldots \\
            &= (-1)^{t_2+t_4+t_5}e^{-i\omega_k(t_1-t_2+t_3-t_4-t_5 \ldots)}.
\end{align}
If we now write
\begin{equation}
 (t_1-t_2+t_3-t_4-t_5 \ldots) = ``\tau"
\end{equation}
we can see that, up to a minus sign, we will obtain expressions of much the 
same form as in the one coin case, but parametrized by $\tau$ instead of 
$t,$ where $\tau$ is defined
\begin{equation}\label{taudef1}
 \tau \in \{\pm t_1 \pm t_2 \pm \ldots \pm t_M\},
\end{equation}
with each $\pm$ independent.  The modulus sign arises naturally when 
the integral expressions for the wavefunction are simplified in a way
analogous to the one coin case, as $\tau$ can be chosen to be positive 
without loss of generality.   When performing the simplification, it is 
also convenient to introduce the integer $j_{\tau},$ which is useful because 
for each $\tau$ we get terms like 
\begin{equation}
 \left(1+\frac{\cos^2(k)}{\sqrt{1+\cos^2(k)}}\right)^{\#(t_is
                                                       \text{ ending in }L)}
 \left(1-\frac{\cos^2(k)}{\sqrt{1+\cos^2(k)}}\right)^{\#(-t_is
                                                       \text{ ending in }L)}
\end{equation}
These are of the form $(a+b)^r(a-b)^s,$ and $j_{\tau}$ counts the 
$(a^2-b^2)^{|r-s|}$ terms in the expansion of that product.  These are of 
the form $(\tfrac{1}{1+\cos^2(k)})^{|r-s|}$ and so they only appear in the 
denominator.  The term raised to the power of $n-2j_{\tau}$ is just the 
two sets of left-overs from that (one multiplied by 
$e^{-i(\omega_k|\tau|+kx)}$ and the other by $e^{i((\omega_k+\pi)|\tau|-kx)})$ 
which have then been combined using the integral simplification identities 
as before.

We can now write down the wavefunction for a particle that started out with 
all the coins in the state $|R\rangle$ and ended in them in (any) one of 
the $\binom{M}{n}$ components with $n$ coins in the state $R:$
\begin{multline}\label{gentemplmR}
 |\psi_{n,(M-n)}(x,t) \rangle = \\
      \sum_{\tau} \frac{1+(-1)^{x+|\tau|}}{2^M}
      \int_{-\pi}^{\pi} \frac{dk}{2\pi} \left(
      \frac{e^{ik(M-n)}\left(\cos(k)+\sqrt{1+\cos^2(k)}\right)^{n-2j_{\tau}}}
      {(1+\cos^2(k))^{M/2}} \right) e^{-i(\omega_k |\tau| + kx)} \;,
\end{multline}
where the sum over $\tau$ indicates a sum over all $2^M$ possible
signs in (\ref{taudef1}).

To write down the wavefunction for a particle that started with 
its coins in any state in the computational basis for the coins, we need 
yet another definition.
We will introduce the notation $\sigma$ to be the $L$-sign weight of $\tau,$ 
which will be the number of minuses in that particular $\tau.$ So for 
\begin{align}
\tau &= t_1+t_2-t_3+t_4-t_5-t_6+t_7, \nonumber\\
\text{initial coin state} &= R\;\quad L\;\quad R\;\quad L\;\quad 
                             L\;\quad L\;\quad R,
\end{align} 
we define $\sigma = 2.$  In other words, for all the coins that started 
out in the state $L,$ count the number of minuses in front of the number 
of turns taken by those coins for that $\tau.$  Ignore any minuses for 
coins that start in the state $R.$  Without loss of generality, $\sigma$ 
can be taken to be in $[0,\lfloor M/2 \rfloor],$ as only $|\tau|$ actually 
matters in the wavefunction. 

Then the wavefunction for a walk starting with its $M$ coins in a state 
in the computational basis of the coins, where there are initially $q$ 
coins in the state $L$ and $M-q$ in the state $R$ will be 
\begin{multline}\label{gentemplmC}
 |\psi_{n,(M-n):\rm{coins}}(x,t) \rangle = 
      \sum_{\tau} \frac{1+(-1)^{x+|\tau|}}{2^M} \\
\times   
      \int_{-\pi}^{\pi} \frac{dk}{2\pi} 
      e^{i\sigma\pi} 
      \left(\sqrt{1 + \cos^2(k)} - \cos(k) \right)^{M-q-\sigma} 
      \left(\sqrt{1 + \cos^2(k)} + \cos(k)\right)^{\sigma} \\
\times
      \left(\frac{e^{-ikn}\left(\cos(k)+\sqrt{1+\cos^2(k)}\right)^
                                                   {n-2j_{\tau}} e^{ikq}}
      {(1+\cos^2(k))^{M/2}} \right) e^{-i(\omega_k |\tau| + kx)}.
\end{multline}

The wavefunction for a general starting state can be obtained from those 
for the computational basis states by forming the corresponding linear 
combination. 

If $\tau$ is an asymptotic parameter: i.e., $\tau \to \infty$ as 
$t \to \infty$ then these integrals may be approximated using the method 
of stationary phase.  If $\tau=0$ or $\tau =$ constant (which can happen 
if the coins are used cyclically: for example if we have two coins, then 
we will get terms where $t_1-t_2 = 0$ or $1$ for all time) then the 
integrals must be performed by some other method.   It is these 
constant-$\tau$ cases that produce the ``stationary'' central spikes seen 
in the simulations for an even number of coins (more on these below). 

If we restrict to the special case where 
$t_1=t_2=\ldots = t_m=t/M$ is an 
integer, then equation \eqref{gentemplmR} simplifies somewhat to 
\begin{multline}
 |\psi_{n,(M-n)}(x,t) \rangle = 
      \sum_{r=0}^{\lfloor M/2 \rfloor} \sum_{j=0}^{\lfloor n/2 \rfloor}
      \binom{M}{r} \frac{1+(-1)^{x+(M-2r)t/M}}{2^M} \times \\ 
      \int_{\pi}^{\pi} \frac{dk}{2\pi} \left(
      \frac{\left(\cos(k)+\sqrt{1+\cos^2(k)}\right)^{n-2j} e^{ik(M-n)}}
      {(1+\cos^2(k))^{M/2}} \right) e^{-i(\omega_k (M-2r)t/M + kx)} \;.
\end{multline}

All of the time-dependent integrals are of the form
\begin{equation}
 I(\alpha,\tau) = \int_{-\pi}^{\pi} \frac{dk}{2\pi} 
                   g(k) e^{i\varphi(k,\alpha_{\tau})|\tau|},
\end{equation}
i.e., these are one-parameter families of generalized Fourier integrals, with 
parameter $\alpha_{\tau} =x/\tau.$  They can therefore be approximated in 
the limit as $\tau \to \infty$ by the method of stationary phase, as used 
in \cite{NayaknV}, and references therein, principally \cite{BleisteinH} 
and \cite{BenderO}.  When each coin is used the same number of times, it can 
be seen that the spikes are equally spaced between $-t/\sqrt{2}$ and 
$t/\sqrt{2}.$  It can also be shown that the height of the peaks scales 
like $1/\sqrt[3]{t}.$

The asymptotic behavior of the time-dependent terms in the above components 
is as follows.  Just as in the one coin case, the integral displays three 
different types of behavior, depending on $\alpha_{\tau},$ and we refer 
readers interested in the details of the analysis to \cite{NayaknV}.  For the  
purposes of calculating the moments of the distribution, we can treat the 
support of the integral as confined to the region 
$\alpha_{\tau} \in (-\frac{1}{\sqrt{2}},\frac{1}{\sqrt{2}})$ to a good 
approximation, and so we will only do the calculations for $\alpha_{\tau}$ in 
this range.  All the integrals have the same form, and differ only by a 
linear rescaling.  

The values of $k$ corresponding to the two stationary points of 
$\varphi(\alpha_{\tau})$ are $\pm k_{\alpha\tau},$ where $k$ is a function 
of $\alpha_{\tau}$, and
\begin{equation}
  \cos(k_{\alpha\tau}) = \frac{-\alpha_{\tau}}{\sqrt{1-\alpha_{\tau}^2}}.
\end{equation}
We also write
\begin{align}
 &\varphi(\alpha_{\tau}) = (\omega_{k_{\alpha\tau}}
                           +\alpha_{\tau} k_{\alpha\tau}) 
 &\omega_{k_{\alpha\tau}} = \arcsin \left(
                        \sqrt{\frac{1-2\alpha_{\tau}^2}
                             {2(1-\alpha_{\tau}^2)}}\right), 
\end{align}
and note that
\begin{align}
 |\omega_{k_{\alpha\tau}}''| &= (1-\alpha_{\tau}^2)\sqrt{1-2\alpha_{\tau}^2} \\
 1 + \frac{\cos(k)}{\sqrt{1+\cos^2(k)}} &= 1-\alpha_{\tau} \\
 \frac{e^{ik}}{\sqrt{1+\cos^2(k)}} &= -\alpha_{\tau} 
                                      + i\sqrt{1-2\alpha_{\tau}^2}.
\end{align}

Using these, we can now write down our asymptotic approximations for the 
time-dependent components.  Let us consider the two-coin walk as an
example.  The time-dependent components 
$\tilde{\psi}(x,t)$ for the two-coin wavefunction are then
\begin{equation}
 |\tilde{\psi}_{RR}(x,t)\rangle = 
 \frac{1+(-1)^{(\alpha+1)t}}{\sqrt{2\pi t(1-\alpha^2)\sqrt{1-2\alpha^2}}}
 (1-\alpha)^2\cos(\varphi(\alpha)t+\pi/4) \;,
\end{equation}
\begin{multline}
 |\tilde{\psi}_{LR}\rangle = |\tilde{\psi}_{RL}\rangle = 
 \frac{1+(-1)^{(\alpha+1)t}}{\sqrt{2\pi t(1-\alpha^2)\sqrt{1-2\alpha^2}}} \\
 \times \left((\alpha(\alpha-1)\cos(\varphi(\alpha)t+\pi/4) 
 + (\alpha-1)\sqrt{1-2\alpha^2}\sin(\varphi(\alpha)t+\pi/4)\right) \;,
\end{multline}
and 
\begin{multline}
  |\tilde{\psi}_{LL}\rangle = 
  \frac{1+(-1)^{(\alpha+1)t}}{\sqrt{2\pi t(1-\alpha^2)\sqrt{1-2\alpha^2}}} \\
 \times \left((3\alpha^2-1)\cos(\varphi(\alpha)t+\pi/4) 
 +2\alpha\sqrt{1-2\alpha^2}\sin(\varphi(\alpha)t+\pi/4)\right) \;,
\end{multline}
where  we have dropped the $\tau$ subscript on $\alpha$ because there is
only one set of asymptotic integrals for this system.  These expressions
can be used to calculate $p(x,t)$ in the long time limit; we plot
this result in figure 3.

More generally, for those components which can be obtained using the 
method of stationary phase, we can write
\begin{multline}\label{tautemp}
 |\psi_{n,(M-n)}(x,t_1,\ldots,t_M) \rangle = 
      \sum_{\tau} \frac{(1+(-1)^{x+|\tau|})(1-\alpha_{\tau})^{(n-2j_{\tau})}
                         (1-\alpha_{\tau}^2)^{(M+j_{\tau}-n/2)}}
      {2^{M-1} \sqrt{2\pi |\tau| (1-\alpha_{\tau}^2)
                      \sqrt{1-2\alpha_{\tau}^2}} } \\
      \times 
      \cos(\varphi(\alpha_{\tau})|\tau| + (M-n)k_{\alpha_{\tau}} + \pi/4),
\end{multline}
where $|\psi_{n,(M-n)}(x,t_1,\ldots,t_M)\rangle$ denotes any of the 
$\binom{M}{n}$ components of the wavefunction which end up with $n$ coins 
in the state $R$ and $(M-n)$ coins in the state $L.$

If $M$ is an even number, the wavefunction will have components 
$\hat{\psi}(x)$ that are central spikes which perform periodic motion for 
a cyclic walk as the walk cycles through the coins.  If the coins are used 
in some other order, these spikes are the final positions of part of the 
wavefunction.  In any event, these central spikes make no contribution to 
the moments of the distribution. 

Let us write the total wavefunction 
$\psi(x,t) = \hat{\psi}(x) + \tilde{\psi}(x,t),$ where $\tilde{\psi}(x,t)$ 
denotes the rest of the components.   For the two-coin 
case, the central spike components are as follows:
\begin{equation}\label{RRspike}
 |\hat{\psi}_{RR}(x) \rangle = \frac{(-1)^{t_1}+(-1)^{t_2}}{8\pi} 
      \int_{-\pi}^{\pi}\frac{e^{-ikx}}{1+\cos^2(k)}dk 
\end{equation}
\begin{multline}\label{LRspike}
 |\hat{\psi}_{LR}(x) \rangle = \frac{(-1)^{t_1}+(-1)^{t_2}}{8\pi} 
      \int_{-\pi}^{\pi}\frac{e^{-ik(x-1)}\cos(k)}{1+\cos^2(k)}dk \quad + \\ 
       \frac{(-1)^{t_1}-(-1)^{t_2}}{8\pi} 
       \int_{-\pi}^{\pi}\frac{e^{-ik(x-1)}}{\sqrt{1+\cos^2(k)}}dk 
\end{multline}
\begin{equation}\label{LLspike}
 |\hat{\psi}_{LL}(x) \rangle = \frac{(-1)^{t_1}+(-1)^{t_2}}{8\pi}
       \int_{-\pi}^{\pi} \frac{e^{-ik(x-2)}}{1+\cos^2(k)}dk. 
\end{equation}
These integrals shrink to zero very rapidly away from 
$x=0,$ but they are still visibly non-zero out as far as $x=\pm 6$ or so. 

These time-independent stationary spike integrals are similar in
form to the time-dependent ones, but this time the method of stationary 
phase isn't helpful.  That asymptotic approximation method can only tell 
us about their behavior as $x \to \infty.$  Since we already know 
that they're independent of time, we can say that these represent 
something that stays in the vicinity of the origin and doesn't go anywhere: 
so we already know they'll tend to zero very rapidly for large $x$.  This 
means that we're only interested in their behavior for a few points in the 
vicinity of the origin.  Since the kernel of the integrand 
oscillates fairly slowly for small values of $x,$ it is practical to 
evaluate these terms numerically.


\bibliographystyle{apsrev}
\bibliography{qwalks}


\end{document}